\documentclass[12pt]{article}

\title{The Pfaff lattice, Matrix integrals and a map from
 Toda to Pfaff}

\author{M. Adler\thanks{Department of Mathematics,
Brandeis University, Waltham, Mass 02454, USA. E-mail:
adler@math.brandeis.edu. The support of a National Science
Foundation grant \# DMS-98-4-50790 is gratefully
acknowledged.}~~~~~~P. van Moerbeke\thanks{Department of
Mathematics, Universit\'e de Louvain, 1348 Louvain-la-Neuve,
Belgium and Brandeis University, Waltham, Mass 02454, USA. E-mail:
vanmoerbeke@geom.ucl.ac.be and @math.brandeis.edu. The  support of
a National Science Foundation grant \# DMS-98-4-50790, a Nato, a
FNRS and a Francqui Foundation grant is gratefully acknowledged.}}

\date{March 8, 1999}

\newcommand{\MAT}[1]{\left(\begin{array}{*#1c}}
\newcommand{\mat}{\end{array}\right)}
\newcommand{\qed}
{%
\mbox{}%
\nolinebreak%
\hfill%
\rule{2mm}{2mm}%
\medbreak%
\par%
}

\newcommand{\sumbis}[2]%
{%

\begin{array}[t]{c}
\sum\\
{\scriptstyle #1}\\
{\scriptstyle #2}
\end{array}

}

\def\boxed#1{\vbox{\hrule\hbox{\vrule$\displaystyle #1$
\vrule}\hrule}}

\newcommand{\pp}{\ldots}

\newcommand{\DR}{{\cal D}}

\newcommand{\BC}{{\Bbb C}}

\newcommand{\BZ}{{\Bbb Z}}
\newcommand{\Bk}{{\Bbb k}}
\newcommand{\Bn}{{\Bbb n}}
\newcommand{\Bu}{{\Bbb u}}

\newcommand{\iy}{\infty}
\newcommand{\pl}{\partial}
\newcommand{\al}{\alpha}
\newcommand{\proof}{\underline{\sl Proof}: }

\newcommand{\HR}{{\cal H}}
\newcommand{\JR}{{\cal J}}

\newcommand{\UR}{{\cal U}}

\newcommand{\TR}{{\cal T}}

\newcommand{\NR}{{\cal N}}

\newcommand{\GR}{{\cal G}}
\newcommand{\vp}{\varphi}
\newcommand{\la}{\langle}
\newcommand{\ra}{\rangle}

\newcommand{\dt}{\delta}

\newcommand{\vr}{\varepsilon}
\newcommand{\sg}{\sigma}
\newcommand{\BR}{{\Bbb R}}

\newcommand{\Lb}{\Lambda}

\def\diag{\mathop{\rm diag}}

\def\be#1\ee{\begin{equation}#1\end{equation}}
\def\bea#1\eea{\begin{eqnarray}#1\end{eqnarray}}
\def\bean#1\eean{\begin{eqnarray*}#1\end{eqnarray*}}

\ifx\undefined\Bbb
        \let\Bbb\bf
\fi

\catcode`\@=11
\def\ps@X{\let\@mkboth\@gobbletwo
        \def\@oddhead{\tt A 
        v  M
        :%
        ~~Pfaff/matrix integrals\hfil March 8, 1999\ \#1\hfil\S\thesection,
p.\thepage
        }
        \def\@oddfoot{\rm\hfil\thepage\hfil}
        \let\@evenhead\@oddhead
        \let\@evenfoot\@oddfoot}
\catcode`@=12
\catcode`\@=11
\let\c@equation=\relax
\newcounter{equation}[subsection]

\catcode`\@=12

\newtheorem{definition}{Definition}[
section]

\newtheorem{theorem}[definition]{Theorem}

\newtheorem{lemma}[definition]{Lemma}
\newtheorem{corollary}[definition]{Corollary}
\newtheorem{proposition}[definition]{Proposition}

\catcode`\@=11
\let\c@equation=\relax
\newcounter{equation}[section]

\catcode`\@=12

\begin{document}
\maketitle






\tableofcontents
\setcounter{section}{-1}

\setcounter{equation}{0}

\newpage
\section{Introduction}
Consider a weight on $\BR$, depending on $t=(t_1,t_2,\ldots)
\in
\BC^{\iy}$,
\be\rho_t(z)dz=e^{\sum t_i
z^i}\rho(z)dz=e^{-V(z)+\sum_1^{\iy}t_i
z^i}dz,~~\mbox{with}~-\frac{\rho'(z)}{\rho(z)}=V'(z)=\frac{g(z)}{f(z)}.\ee

{\bf Hermitean matrix integrals} ({\sl revisited})  This weight
leads to a $t$-dependent moment matrix
 $$
 m_n(t)=(\mu_{k+\ell}(t))_{0
 \leq k,\ell\leq n-1}
 =\left( \int_{\BR}z^{k+\ell}\rho_t(z) dz
 \right)_{0
 \leq k,\ell\leq n-1},
$$
with the semi-infinite moment matrix $m_{\iy}$, satisfying the
commuting equations
\be\pl m_{\iy} / \pl t_k=\Lb^k m_{\iy}= m_{\iy}\Lb^k,\ee
where $\Lb$ is the customary shift matrix. Considering the lower-
and an upper-triangular matrix Borel decomposition
\be
m_{\iy}=S^{-1} S^{\top -1},
\ee
which is determined by the following $t$-dependent matrix
 integrals\footnote{We set vol$( \UR(n))=1$ for all $n$.} ($n\geq 0$)
\be
\tau_n(t):=\int_{\HR_n}e^{Tr (- V(X)+\sum t_i X^i)}dX
=\det m_n, ~~\mbox{and}~~\tau_0=1,
\ee
with Haar measure $dX$ on the ensemble
$
{\cal H}_n=\{n\times n\mbox{\,\,Hermitean matrices}\}.
$
As is well known, the integral (0.4) is a solution to the following
two systems,
\newline\noindent{\em (i) the KP-hierarchy}
 \be
\left(p_{k+4}(\tilde\pl)-\frac{1}{2}\frac{\pl^2}{\pl
t_1\pl t_{k+3}}\right)\tau_{n}\circ\tau_{n}=0,
 ~~\mbox{for}~~ k,n=0,1,2,\dots.
\ee
\newline\noindent{\em (ii) the Toda lattice}; i.e., the tridiagonal
matrix
 \be
L(t):= S\Lb S^{ -1}=\pmatrix{
 \frac{\pl}{\pl t_1}\log \frac{\tau_1}{\tau_0} &
 \left(\frac{\tau_{0}\tau_{2}}{\tau_{1}^2}\right)^{1/2}& 0 \cr
\left(\frac{\tau_{0}\tau_{2}}{\tau_{1}^2}\right)^{1/2}&
\frac{\pl}{\pl t_1}\log \frac{\tau_2}{\tau_1} &
\left(\frac{\tau_{1}\tau_{3}}{\tau_{2}^2}\right)^{1/2}\cr
0&\left(\frac{\tau_{1}\tau_{3}}{\tau_{2}^2}\right)^{1/2}&
 \frac{\pl}{\pl t_1}\log \frac{\tau_3}{\tau_2} & \ddots \cr
&  ~~~~\ddots &~~~~~~~~~\ddots }\ee satisfies the following
commuting Toda equations
$$
\frac{\pl L}{\pl t_n}=\left[ \frac{1}{2}(L^n)_{sk},L \right],
$$
where $(A)_{sk}$ denotes the skew-part of the matrix $A$ for the
Lie algebra splitting into skew and lower-triangular matrices.
Moreover, the following $t$-dependent polynomials in $z$,
 are defined
by the $S$-matrix obtained from the Borel decomposition (0.3);
 it is also given, on the one hand, by a classic
 determinantal formula,
and on the other hand, in terms of the functions $\tau_n(t)$:
\bean
p_n(t,z):&=&(S(t)\chi(z))_n=\frac{1}{\sqrt{\tau_n\tau_{n+1}}}\det
\left(\begin{array}{ccc|c}
 & & &1\\
 & & &z \\
 &m_n(t)& &\vdots\\
 & & & \\
\hline
\mu_{n,0}(t)&\pp&\mu_{n,n-1}(t)&z^n
\end{array}\right)\\
&=& z^n \frac{\tau_n(t-[z^{-1}])}{\sqrt{\tau_n \tau_{n+1}}}
.\eean
The $p_n$'s are orthonormal with respect to the (symmetric)
inner-product $\la,\ra_{sy}$, defined by $\la
z^i,z^j\ra_{sy}=\mu_{ij}$, which is a restatement of the Borel
decomposition (0.3). The vector $p(t,z)=(p_n(t,z))_{n\geq 0}$ is an
eigenvector of the matrix $L(t)$ in (0.6):
$$L(t)p(t,z)=zp(t,z).$$

\noindent{\bf Symmetric and symplectic matrix integrals.}
~ Instead consider the following skew-symmetric matrix
$
 m_{\iy}=(\mu_{ij})_{i,j\geq 0}$ of moments\footnote{$\vr(x)=1$, for
$x\geq 0$ and $=-1$, for $x<0$ and $\{f,g\}=f'g-fg'$.}
\be\mu^{(1)}_{ij}(t)=\int\!\int_{\BR^2}x^iy^j\vr(x-y)\rho_t(x)\rho_t(y)dx dy
~~~\mbox{or}~~~
\mu^{(2)}_{ij}(t)=\int_{\BR}
 \{y^i,y^j\}\rho_t(y)^2 dy,
 \ee
both satisfying the equations
\be
\frac{\pl m_{\iy}}{\pl t_i}=\Lb^im_{\iy}+m_{\iy}\Lb^{\top i}.
\ee
In this paper, we consider symmetric matrix
 integrals\footnote{where again we set the volume of the
 orthogonal and symplectic groups equal to $1$.}
 \be
\tau^{(1)}_{2n}(t):= \frac{1}{(2n)!}\int_{{\cal S}_{2n}}
 e^{Tr~(- V(X)+\sum_1^{\iy} t_i X^i )} dX=pf(m^{(1)}_{2n}),
\ee
and symplectic matrix integrals
\be
 \tau^{(1)}_{2n}(t):= \frac{1}{n!}\int_{{\cal T}_{2n}}
 e^{2Tr~(- V(X)+\sum_1^{\iy} t_i X^i )} dX=pf(m^{(2)}_{2n}),
 \ee
both expressed in terms of the Pfaffian of the ``moment"
 matrix $m^{(i)}_{\iy},$
where
\newline (1) in the first case, $dX$ denotes Haar measure
on the space $S_{2n}$ of symmetric matrices and,
\newline (2) in the
second case, $dX$ denotes Haar measure
 on the $2n
\times 2n$ matrix realization $
\TR_{2n}$ of the space of self-dual $n\times n$ Hermitean matrices, with quaternionic
entries.
\newline Since $m_{\iy}$ is skew-symmetric, the Borel decomposition of
 $m_{\iy}$ will require the interjection of a skew-symmetric
 matrix $J,$ used throughout this paper,
\be
J=\left(
\begin{array}{cc@{}c@{}cc}
\ddots &&&& \\
 &\boxed{\begin{array}{cc} 0 & 1 \\ -1 & 0 \end{array}} &&0& \\
 && \boxed{\begin{array}{cc} 0 & 1 \\ -1 & 0 \end{array}} &&\\
 &0&& \boxed{\begin{array}{cc} 0 & 1 \\ -1 & 0 \end{array}} & \\
 &&&& \ddots
 \end{array}
 \right)~~\mbox{with}~J^2=-I
 \ee
 and the order 2 involution on the space $\DR$ of
 infinite matrices
 \be
 \JR:\DR\longrightarrow
\DR: a\longmapsto\JR(a):=Ja^{\top}J.
\ee
The skew-Borel decomposition
 \be
m_{\iy}(t)=Q^{-1}(t)JQ^{-1
\top}(t)
,\ee
 can entirely be expressed in terms of the integrals $\tau_{2n}(t)$,
  (0.9) and (0.10) corresponding respectively to the first and second
   moment matrix (0.7). They satisfy both
\newline\noindent{\em (i) the Pfaffian KP-hierarchy} for
$k,n=0,1,2,\dots$\,,
\be
\left(p_{k+4}(\tilde\pl)-\frac{1}{2}\frac{\pl^2}{\pl
t_1\pl t_{k+3}}\right)\tau_{2n}\circ\tau_{2n}=p_k(\tilde
\pl)~\tau_{2n+2}\circ\tau_{2n-2},
\ee
\newline\noindent{\em (ii) the Pfaff lattice}; i.e., the matrix,
constructed by dressing up $\Lb$ with $Q$ and which this time is
full below the main diagonal,
$$ L =Q\Lb Q^{-1}=
 h^{-1/2} \pmatrix{
 \hat L_{00}&\hat L_{01}&0&0&\cr
\hat L_{10}&\hat L_{11}&\hat L_{12}&0&\cr *&\hat L_{21} &\hat
L_{22}&\hat L_{23}&\cr
 *&*
&\hat L_{32}&\hat L_{33}&\dots\cr  & & &\vdots& &
\cr} h^{1/2},
$$
satisfies the Hamiltonian commuting equations
\be
\frac{\pl L}{\pl
t_i}=
\left[\left((L^i)_++\JR ((L^i)_+)\right)+\frac{1}{2}
\left((L^i)_0+ \JR ((L^i)_0) \right),L\right],
\ee
with the entries $\hat L_{ij}$ and the entries of $h$, being
$2\times 2$ matrices
$$
h=\mbox{diag}(h_0I_2,h_2I_2,h_4I_2,\dots),
~h_{2n}=\tau_{2n+2}/\tau_{2n},
$$
and ($ {}^.=\frac{\pl}{\pl t_1}$)
$$
\hat L_{nn}:=\pmatrix{-(\log \tau_{2n})^. & &1 \cr \cr
 -\frac{S_2(\tilde \pl)\tau_{2n}}{\tau_{2n}}
-\frac{S_2(-\tilde \pl)\tau_{2n+2}}{\tau_{2n+2}} & &
 (\log \tau_{2n+2})^.\cr}~~
 ~~~~~~\hat L_{n,n+1}:=\pmatrix{0& 0 \cr
 1 & 0\cr}
$$

\vspace{0.4cm}

\be
\hat L_{n+1,n}:=\pmatrix{*&(\log \tau_{2n+2})^{..}&  \cr
 * &*\cr}.
 \ee
The following $t$-dependent polynomials $q_n(t,z)=(S\chi(z))_n$ in
$z$, defined by the $S$ matrix of the skew-Borel
 decomposition (0.13), have determinantal and
 Pfaffian $\tau$-function expression, in analogy with
  the Hermitean case:
\bean
q_{2n}(t,z)&=&\frac{1}{\sqrt{\tau_{2n}(t)\tau_{2n+2}(t)}}pf
\left(\begin{array}{ccc|c}
 & & &1\\
 & & &z \\
 &m_{2n+1}(t)& &\vdots\\
 & & &z^{2n} \\
\hline
-1&\pp&-z^{2n}&0
\end{array}\right)\\
&=& z^{2n} \frac{\tau_{2n}(t-[z^{-1}])}
 {\sqrt{\tau_{2n}(t) \tau_{2n+2}(t)}}
\eean
and
\bea
q_{2n+1}(t,z)&=&\frac{1}{\sqrt{\tau_{2n}\tau_{2n+2}}}pf
\left(\begin{array}{ccc|cc}
 & & & 1&\mu_{0,2n+1}\\
& & & z &\mu_{1,2n+1} \\
 &m_{2n}(t) & & \vdots&\vdots\\
& & &z^{2n} &\mu_{2n,2n+1} \\
\hline
-1&-z&\pp&0 &-z^{2n+1}\\
-\mu_{0,2n+1}&-\mu_{1,2n+1} &\pp&z^{2n+1}&0
\end{array}\right)\nonumber\\
&=&\frac{1}{\sqrt{\tau_{2n}(t)\tau_{2n+2}(t)}}
\left( z+\frac{\pl}{\pl t_1} \right)
\sqrt{\tau_{2n}(t)\tau_{2n+2}(t)}~q_{2n}(t,z)\nonumber\\
&=& z^{2n} \frac{ (z+\frac{\pl}{\pl t_1})\tau_{2n}(t-[z^{-1}])}
 {\sqrt{\tau_{2n}(t) \tau_{2n+2}(t)}}
\eea
form skew-orthonormal sequences with respect to the skew
inner-product
 $\la,\ra_{sk}$, defined by $\la y^i,z^j\ra_{sk}=\mu_{ij}$,
 namely, we have the following restatement of the
 skew-Borel dcomposition (0.13):
\be
(\la  q_i,  q_j\ra)_{0\leq i,j<\iy}=J.
\ee
Finally, the vector $q(z)=(q_n(z))_{n\geq 0}$ forms a eigenvector
for the matrix L:
\be
L(t)q(t,z)=zq(t,z).
\ee

In section 2, we show how a general skew-symmetric infinite matrix
 flowing via (0.8) and its skew-Borel decomposition
 (0.13), lead to wave vectors $\Psi$,
 satisfying bilinear relations and differential equations.
 Section 3 deals with the existence, in the above general
 setting, of a so-called  Pfaffian
 $\tau$-function, satisfying bilinear equations and a KP-
 type hierarchy. In \cite{ASV}, these results were obtained,
 by embedding the system in 2-Toda theory, while in this paper,
 they are obtained in an intrinsic fashion.

For $k=0$, the KP-like equation (0.14) has already appeared in the
context of the charged BKP hierarchy, studied by V. Kac and van de
Leur \cite{KvdL}; the precise relationship between the charged BKP
hierarchy of Kac and van de Leur and the Pfaff Lattice, introduced
here, deserves further investigation. See the recent paper of
van de Leur \cite{vdL}.

\vspace{1cm}

\noindent {\bf A remarkable map from Toda to
 Pfaff lattice}:
Remembering the notation (0.1), we act with
 the $z$-operator,
\be
\Bn_t:=\sqrt{\frac{f}{\rho_t}}\frac{d}{dz}
\sqrt{f\rho_t}
=e^{-\frac{1}{2}\sum t_k
z^k}\left(\frac{d}{dz}f(z)-\frac{f'+g}{2}(z)\right)
e^{\frac{1}{2}\sum t_k z^k}
\ee
on the $t$-dependent orthonormal
  polynomials $p_n(t,z)$ in $z$; in \cite{AvM2}, we
  showed that the matrix $\NR$ defined by
  \be
  \Bn_t p(t,z)= \left(f(L)M-\frac{f'+g}{2}(L)\right)p(t,z)
  =: \NR p(t,z)
  \ee
is skew-symmetric. The $t$-dependent matrix $\NR$ is expressed in
terms of $L$ and a new matrix $M$, defined by
 \be
 zp=Lp~~~\mbox{and}~~~
 e^{-\frac{1}{2}\sum t_k z^k}\frac{d}{dz}
 e^{\frac{1}{2}\sum t_k z^k}p=Mp.
 \ee
Consider now the skew-Borel decomposition of $\NR(2t)$ and its
inverse\footnote{See the appendix, section 9.} $\NR(2t)^{- 1}$, in
terms of lower-triangular matrices $O_{(+) }(t)$ and $O_{(-)}(t)$
respectively\footnote{The upper-signs (respectively, lower-signs)
correspond to each other throughout this section.}
:
\be
\NR(2t)^{\pm 1}=-O_{(\pm )}^{-1}(t)JO_{(\pm) }^{\top -1}(t).
\ee
Then, the lower-triangular matrices $O_{(\pm)}(t)$ map {\em
orthonormal} into {\em skew-orthonormal} polynomials, and the
tridiagonal $L$-matrix into an $\tilde L$-matrix:
\bea
p_n(t,z)~~~~~~&\longmapsto& q^{(\pm)}_n(t,z)=O_{(\pm)}(t)p_n(t,z)
 \nonumber\\
L(t)~~~~~~~~ ~&\longmapsto& \tilde
L(t)=O_{(\pm)}(t)L(2t)O_{(\pm)}(t)^{-1} .\\
\mbox{({\bf Toda Lattice})}& & ~~~\mbox{({\bf Pfaff Lattice})}
 \nonumber
\eea
It also maps the weight into a new weight
$$\rho(z)=e^{-V(z)}\longmapsto \tilde
 \rho_{\pm}(z)=e^{-\tilde V(z)}:=e^{-\frac{1}{2}(V(z)\mp\log
f(z))},$$
 and the corresponding string of $\tau$-functions into
 a new string of pfaffian $\tau$-functions: (remember $V_t(z)=V(z)-\sum_1^{\iy} t_i z^i$)
$$
\tau_k(t)=\int_{{\cal H}_{k}}
 e^{Tr~(- V_t(X))} dX\longmapsto \left\{\begin{array}{l}
 \displaystyle
\tau^{(+)}_{2n}(t):= \int_{{\cal T}_{2n}}
 e^{Tr~2(- \tilde V_t(X))} dX ,~( \beta=4)
  \\[10pt]
 \displaystyle
  \tau^{(-)}_{2n}(t):= \int_{{\cal S}_{2n}}
 e^{Tr~(- \tilde V_t(X) )} dX~ ~( \beta=1) .
\end{array}
\right.
$$
For the {\bf classical orthogonal polynomials} $p_n(z)$, we have
shown in
 \cite{AvM2}, that $\NR(0)$ is not only skew-symmetric, but also
 tridiagonal; i.e.,
\be
L=\left[\begin{array}{ccccc}
b_0&a_0& & & \\
a_0&b_1&a_1& & \\
 &a_1&b_2&\ddots&\\
 & &\ddots&\ddots&
\end{array}
\right],\quad -\NR=\left[\begin{array}{ccccc}
0&c_0& & & \\
-c_0&0&c_1& & \\
 &-c_1&0&\ddots&\\
 & &\ddots& &
\end{array}\right] .
\ee
In section 6 and 7, we show that the maps $O_{(-)}$ and $O_{(+)}$
only involves three steps, in the following sense:
\bea
 q^{(-)}_{2n}(0,z)&=&\sqrt{\frac{c_{2n}}{a_{2n}}}p_{2n}(0,z)\nonumber\\
 q^{(-)}_{2n+1}(0,z)&=&\sqrt{\frac{a_{2n}}{c_{2n}}}\nonumber\\
&&
\left(-c_{2n-1}p_{2n-1}(0,z)+\frac{c_{2n}}{a_{2n}}(\sum_0^{2n}b_i)
 p_{2n}(0,z)+c_{2n}p_{2n+1}(0,z)\right)\nonumber\\
&&\hspace{8cm} (\beta=1)
\eea
\bea
 p_{2n}(0,z)&=&
-c_{2n-1}\sqrt{\frac{a_{2n-2}}{c_{2n-2}}}q^{(+)}_{2n-2}(0,z)
+\sqrt{a_{2n}c_{2n}}~q^{(+)}_{2n}(0,z)\nonumber\\
 p_{2n+1}(0,z)&=&
 -c_{2n}\sqrt{\frac{a_{2n-2}}{c_{2n-2}}}
 q^{(+)}_{2n-2}(0,z)-(\sum_0^{2n} b_i)\sqrt{\frac{c_{2n}}{a_{2n}}}
 q^{(+)}_{2n}(0,z)+\sqrt{\frac{c_{2n}}{a_{2n}}}q^{(+)}_{2n+1}(0,z),\nonumber\\
 &&\hspace{8cm} (\beta=4)
 \eea

The abstract map $O_{(-)}$ for $t=0$ appears already in the
work of E. Br\'ezin and H. Neuberger \cite{BN}. This
has been applied in \cite{AFNV} to a problem
in the theory of random matrices.

\section{Splitting theorems, as applied to the Toda
and Pfaff Lattices}

In this section, we show how each of the equations
\be
\frac{\pl m_{\iy}}{\pl t_i}=\Lb^im_{\iy}~~~\mbox{and}~~~
\frac{\pl m_{\iy}}{\pl t_i}=\Lb^im_{\iy}+m_{\iy}\Lb^{\top i},
\ee
lead to commuting Hamiltonian vector fields related to a Lie
algebra splitting. First recall the splitting theorem, due to
Adler-Kostant-Symes \cite{AvM1} and the R-version to
 Reiman and
Semenov-Tian-Shansky \cite{RS}. The R-version allows for more
general initial conditions.

 \begin{proposition}  Let ${\bf g} = {\bf k} + {\bf n}$ be a
 (vector space) direct sum of a Lie algebra  ${\bf g}$ in terms of Lie
 subalgebras  ${\bf k}$ and  ${\bf n}$, with ${\bf g}$ paired with
 itself via a non-degenerate {\rm ad}-invariant inner
 product\footnote{$\la{\rm Ad}_gX;Y\ra =\la X,{\rm Ad}_{g-1}Y\ra$,
 $g\in G$, and thus $\la [z,x],y\ra =\la x,-[z,y]\ra.$} $\la\,
 ,\,\ra$; this in turn induces a decomposition ${\bf g} = {\bf
 k}^{\bot} + {\bf n}^{\bot}$ and isomorphisms  ${\bf g}\simeq {\bf
 g}^*$, ${\bf k}^{\bot}\simeq {\bf n}^*$, ${\bf n}^{\bot}\simeq {\bf
 k}^*$.   $\pi_{{\bf k}}$ and  $\pi_{{\bf n}}$ are
 projective onto  ${\bf k}$ and ${\bf n}$ respectively. Let $\GR, ~\GR_{\Bk}$
 and $ \GR_{\Bn}$ be the groups associated with the Lie algebras  ${\bf g},
 {\bf k}$ and ${\bf n}$. Let ${\cal I}({\bf g})$
 be the {\rm Ad}$^* \simeq$ {\rm Ad}-invariant functions on  ${\bf
 g}^* \simeq {\bf g}$.

 {\bf (i)} Then, given an element
 $$
 \vr \in   {\bf g} :
 [\vr,{\bf k}] \subset {\bf k}^{\bot} \mbox{\,\, and\, \,} [\vr,{\bf
 n}]  \subset {\bf n}^{\bot},
 $$
 the functions
 \be
 \vp(\vr +\xi')|_{{\bf k}^{\bot}}\mbox{\, with \,}
 \vp\in {\cal I}({\bf g})~ \mbox{and}~ \xi' \in
 {\bf k}^{\bot}, \ee
 respectively Poisson commute for the
 respective Kostant-Kirillov  symplectic structures of $n^* \simeq
 {\bf k}^{\bot}$; the associated
 Hamiltonian flows are expressed in terms of the Lax
 pairs\footnote{$\nabla\vp$ is defined as the element in ${\bf g}^*$
 such that $d\vp(\xi)=\la\nabla\vp,d\xi\ra$, $\xi\in{\bf g}$.}
 \be
 \dot\xi = [-\pi_{{\bf k}}\nabla\vp(\xi),\xi]=[\pi_{{\bf
 n}}\nabla\vp(\xi),\xi]\mbox{\,for\,}
 \xi \equiv\vr+\xi',\xi'\in{\bf k}^{\bot}
 \ee

{\bf (ii)} The splitting also leads to a second Lie algebra  ${\bf
g}_R$, derived
 from ${\bf g}$, such that ${\bf g}^*_R\simeq {\bf g}_R$, namely:
 \be {\bf g}_R : [x,y]_R = \frac{1}{2}[Rx,y] +
 \frac{1}{2}[x,Ry] = [\pi_{{\bf k}}x,\pi_{{\bf k}}y] - [\pi_{{\bf
 n}}x,\pi_{{\bf n}}y], \ee
 with  $R = \pi_{{\bf
 k}}-\pi_{{\bf n}}$. The functions
$$\vp(\xi)|_{{\bf g}_R}
 \mbox{\, with \,}\vp\in {\cal I}({\bf g})~ \mbox{and}~
 \xi \in  {\bf g}_R $$ respectively Poisson commute for the
 respective Kostant-Kirillov  symplectic structures
 of ${\bf g}^*_R\simeq{\bf g}_R$, with the same associated
 (Hamiltonian) Lax pairs
 \be
 \dot\xi = [-\pi_{{\bf k}}\nabla\vp(\xi),\xi]=[\pi_{{\bf
 n}}\nabla\vp(\xi),\xi]~~\mbox{\,for\,}~~
 \xi\in{\bf g}_R.
 \ee
 Each of the equations (1.3) and (1.5) has the same solution expressible in two different
ways\footnote{naively written Ad$_{K(t)}\xi_0=K(t)\xi_0K(t)^{-1}$,
Ad$_{S^{-1}}\xi_0=S^{-1}(t)\xi_0S(t)$.}:
\be
\xi(t)={\rm Ad}_{K(t)}\xi_0={\rm Ad}_{S^{-1}(t)}\xi_0,
\ee
with\footnote{with regard to the group factorization
$A=\pi_{\GR_{\Bk}}A~
\pi_{\GR_{\Bn}}A$.}
$$
K(t) = \pi_{\GR_{\Bk}} e^{t \nabla\vp(\xi_0)},\quad
 \mbox{and}\quad
S(t)
=
 \pi_{\GR_{\Bn}} e^{t \nabla\vp(\xi_0)}.
$$
\end{proposition}

\noindent {\sl \underline{Example 1}}:
{\bf The standard Toda lattice and the equations
$\displaystyle{\frac{\pl m}{\pl t_i}}=\Lambda^im $ for the H\"ankel
matrix $m_{\iy}$}. Since, in particular, the matrix $m_{\iy}$ is
symmetric,
 the Borel decomposition into
lower- times upper-triangular matrix must be done with the same
lower-triangular matrix $S$:
\be
m_{\iy}=S^{-1} S^{\top -1}.
\ee
In turn, the matrix $S$ defines a wave vector $\Psi$, and
 operators\footnote{In the formulas below $\chi(z)=(z^0,z,z^2,\dots)$ and
 $\pl$ is the matrix such that $\frac{d}{dz}\chi(z)
 =\pl \chi(z)$. } $L$ and $M$, the same as the
 ones defined in (0.22),
\be
\Psi(t,z):=e^{\frac{1}{2} \sum_1^{\iy}t_i z^i}S\chi,
~~~L:=S\Lb S^{-1},~~~ M:=S(\pl +\frac{1}{2}\sum_1^{\iy}it_i
\Lb^{i-1})S^{-1}, \ee
 satisfying the following well-known equations\footnote
 {where the $()_{sk}$ and $()_{bo}$ refers to the
 skew-part and the lower-triangular (Borel) part respectively;
 i.e., projection onto $\Bk$ and $\Bn$ respectively.}:
\bea
L\Psi=z\Psi~~&\mbox{ }&~~M\Psi=\frac{\pl}{\pl z}\Psi,\quad
\mbox{with}\quad [L,M]=1,\nonumber\\
\frac{\pl S}{\pl
t_n}=-\frac{1}{2}{(L^n)}_{bo} S ~~&\mbox{ }&~~\frac{\pl
\Psi}{\pl t_n}=\frac{1}{2}{(L^n)}_{sk}  \Psi
\nonumber\\
 \frac{\pl L}{\pl t_n}=\frac{1}{2}[(L^n)_{sk},L]~~
 &\mbox{ }&~~
 \frac{\pl M}{\pl t_n}=\frac{1}{2}[(L^n)_{sk},M].
\eea
 The wave vector $\Psi$ can then be expressed in terms of a sequence
 of $\tau$-functions $\tau_n(t)=\det m_n(t)$, but also
 has the simple expression in terms of orthonormal
 polynomials, with respect to the moment matrix
 $m_{\iy}$:
 \bea
 \Psi(t,z)&=& e^{\frac{1}{2}\sum t_iz^i}
\left(z^n\frac{\tau_n(t-[z^{-1}])}
{\sqrt{\tau_n(t)\tau_{n+1}(t)}}\right)
_{n\geq 0}
\nonumber\\
 &=& e^{\frac{1}{2}\sum t_iz^i}\left(p_n(t,z)\right)_{n\geq 0}
.
 \eea

The vector fields (1.9) on $L$ are commuting Hamiltonian vector
fields, in view of the Adler-Kostant-Symes splitting theorem
(version (i)),
\be
\frac{\pl  L}{\pl t_i}=[-\pi_{\Bk}\nabla \HR_i,L]=
[\pi_{\Bn}\nabla \HR_i,L],~~~\HR_i=\frac{tr L^{i+1}}{i+1}, ~~~
\nabla \HR_i=L^i,
\ee
with
\be
L=\Lb^{\top} a+b  +a \Lb, ~~\mbox{ $a$ and $b$ diagonal matrices}
\ee
 for the splitting of the Lie algebra of semi-infinite
 matrices
\bea
\DR=gl_{\iy}&=&\Bk+\Bn:=\{\mbox {skew-symmetric }\}+
\{ \mbox {lower-triangular} \}  \nonumber\\
&=&\Bk^{\bot}+\Bn^{\bot}:=\{\mbox {symmetric }\}+
\{ \mbox {strictly upper-triangular} \} , \nonumber\\
\eea
with the form (1.12) of $L$ being preserved in time.
 Note that the solution (1.6) to (1.5) in the AKS theorem is
 nothing but the factorization of $m_{\iy}$ followed
 by the dressing up of $\Lb$.

\noindent {\sl \underline{Example 2}}:
{\bf The Pfaff lattice and the equations $\displaystyle{\frac{\pl
m}{\pl t_i}}=\Lambda^im+m\Lambda^{\top^i}.$}

Throughout this paper the Lie algebra $\DR = gl_{\iy}$ of
semi-infinite matrices is viewed as composed of $2\times 2$ blocks.
It admits the natural decomposition into subalgebras:
\be
\DR=\DR_-\oplus\DR_0\oplus\DR_+=\DR_-\oplus\DR_0^-\oplus\DR_0^+\oplus\DR_+
\ee
where $\DR_0$ has $2\times 2$ blocks along the diagonal with zeroes
everywhere else and where $\DR_+$ (resp. $\DR_-$) is the subalgebra
of upper-triangular (resp. lower-triangular) matrices with $2
\times 2$ zero matrices along $\DR_0$ and
zero below (resp. above). As pointed out in (1.14), $\DR_0$ can
further be decomposed into two Lie subalgebras:
\bea
\DR_0^-&=&\{\mbox{all $2\times 2$ blocks $\in \DR_0$
are proportional to Id}   \}\nonumber\\
\DR_0^+&=&\{\mbox{all $2\times 2$ blocks $\in \DR_0$
have trace $0$ }  \}.
\eea
Remember from (0.10) and (0.11) in the introduction, the matrix $J$
and the associated Lie algebra order 2 involution $\JR$. The
splitting into two Lie subalgebras\footnote{Note $\Bn$ is the fixed
point set of $\JR$.}
\be
\DR= \Bk+\Bn,
\ee
with
 \bea \Bk&=&\DR_-+\DR_0^-\nonumber\\
 &=&~\mbox{algebra of}~
\left(
\begin{array}{c@{}c@{}cc}
\ddots &&&0 \\
 & \boxed{\begin{array}{cc}
 Q_{2n,2n} & 0 \\ 0 & Q_{2n,2n} \end{array}} &&\\
 &*& \boxed{\begin{array}{cc}
  Q_{2n+2,2n+2} & 0 \\ 0 & Q_{2n+2,2n+2} \end{array}} & \\
 &&& \ddots
 \end{array}
 \right)
 ~~\nonumber
\\
{\Bbb n}&=& \{a\in \DR, ~\mbox{such that}~ \JR a=a\}=\{b+\JR b,~b
\in \DR  \}=\mbox{sp}(\iy),
\eea
with corresponding Lie groups\footnote{$\GR_{\Bk}$ is the group of
invertible elements in $\Bk$, i.e., invertible lower-triangular
matrices, with non-zero $2\times 2$ blocks proportional to Id along
the diagonal.} $\GR_{\Bk}$ and $\GR_{\Bn}=Sp(\iy)$, will play a
crucial role here. Let $\pi_{\Bk}$ and $\pi_{\Bn}$ be the
projections onto $\Bk$ and $\Bn$. Notice that $\Bn=$sp$(\iy)$ and
$\GR_{\Bn}=Sp(\iy)$ stand for the infinite rank affine symplectic
algebra and group; e.g. see \cite{Kac}. Any element $a\in\DR$
decomposes uniquely into its projections onto $\Bk$ and $\Bn$, as
follows:
\bea
a&=&\pi_{\Bk}a+\pi_{\Bn}a\nonumber\\ &=&\left\{\left(a_--\JR
a_+\right)+\frac{1}{2}\left(a_0-\JR a_0
\right)\right\}+\left\{\left(a_++\JR a_+\right)+\frac{1}{2}\left(a_0+\JR
a_0
\right)\right\}.\nonumber\\
\eea
The following splitting, with
$$
\Bk_+=\DR_++\DR_0^-~~\mbox{and}~~\Bn_+=\Bn,
$$
 will also be used in section 2; the projections take on
 the following form,
\bea
a&=&\pi_{\Bk_+}a+\pi_{\Bn_+}a\nonumber\\
 &=&\left\{\left(a_+-\JR
a_-\right)+\frac{1}{2}\left(a_0-\JR a_0
\right)\right\}+\left\{\left(a_-+\JR a_-\right)+
\frac{1}{2}\left(a_0+\JR a_0
\right)\right\}.\nonumber\\
\eea
Note $\JR$ intertwines $\pi_{\Bk}$ and $\pi_{\Bk_+}$:
\be
\JR \pi_{\Bk}=\pi_{\Bk_+}   \JR.
\ee

\medbreak

For a skew-symmetric semi-infinite matrix $m_{\iy}$, the skew-Borel
decomposition \be m_{\iy}:=Q^{-1}J Q^{-1\top}\mbox{ with }Q\in
\GR_{\Bk},\ee is unique, as was shown in \cite{AHV}. Here we may assume
 $m_{\iy}$ to be bi-infinite, as long as the factorization
  (1.21)
 is unique, upon imposing a suitable normalization. Then we use $Q$ to dress up $\Lb$:
 $$
 L=Q\Lb Q^{-1}.
 $$
Then letting $m_{\iy}$ run according to the equations
$
\pl
m / \pl t_i=\Lambda^i m+m\Lambda^{\top^i},
$
we show in the next proposition and corollary that $L$ evolves
according to a system of commuting equations, which by virtue of
the AKS theorem are Hamiltonian vector fields; for details, see
\cite{AHV}.

\begin{proposition} For the matrices
$$
m_{\iy}:=Q^{-1}J Q^{-1\top} \quad\mbox{and}\quad L:=Q\Lb
Q^{-1}\quad ,
\mbox{ with }Q\in \GR_{\Bk},
$$
the following three statements are equivalent\medbreak\indent (i) $\displaystyle{\frac{\pl Q}{\pl
t_i}Q^{-1}=-\pi_{\Bk}L^i}$
\medbreak\indent (ii) $L^i+\displaystyle{\frac{\pl Q}{\pl
t_i}}Q^{-1}\in\Bn$\medbreak
 \indent (iii) $\displaystyle{\frac{\pl m}{\pl
t_i}}=\Lambda^im+m\Lambda^{\top^i}.$
\medbreak\noindent Whenever the vector fields on $Q$ or $m$
satisfy (i), (ii) or (iii), then the matrix $L=Q\Lb Q^{-1}$ is a
solution of the AKS-Lax pair
$$
\frac{\pl L}{\pl t_i}=[-\pi_{\Bk}L^i,L]=
 \left[\pi_{{\bf n}} L^i,L    \right].
$$

\end{proposition}

\proof Written out and using (1.18), proposition 1.2 amounts to showing the
equivalence of the three formulas:
\medbreak
\noindent(I) $\displaystyle{\frac{\pl Q}{\pl t_i}Q^{-1}+\left((L^i)_--J(L^i_+)^{\top}J\right)+\frac{1}{2}
\left((L^i)_0-J((L^i)_0)^{\top}J\right)=0}$

\medbreak

\noindent(II) $\displaystyle{\left(L^i+\frac{\pl Q}{\pl t_i}Q^{-1}\right)-J
\left(L^i+\frac{\pl Q}{\pl t_i}Q^{-1}\right)^{\top}J=0}$

\medbreak

\noindent(III) $\Lambda^im+m\Lambda^{\top}-\displaystyle{\frac{\pl m}{\pl t_i}}=0$.

\medbreak

\noindent The point is to show that
$$
(\mbox{I})_+=0,~~~~(\mbox{I})_-=(\mbox{II})_-
=-J~(\mbox{II}_+)^{\top} ~J,~~~~
(\mbox{I})_0=\frac{1}{2}\left(\mbox{II}\right)_0,
$$
\be
Q^{-1}(\mbox{II})JQ^{-1^{\top}}
=(\mbox{III}).
\ee
The reader will find the details of this proof in \cite{AHV}.
\qed

\section{Wave functions and their bilinear equations for
the Pfaff Lattice}

\setcounter{equation}{-1}

Consider the commuting vector fields \be \pl m_{\iy}/ \pl
t_i=\Lb^im_{\iy}+m_{\iy}\Lb^{\top i} \ee on the skew-symmetric
matrix $m_{\iy}(t)$ and  the skew-Borel decomposition
\be
m_{\iy}(t)=Q^{-1}(t)J\,Q^{\top -1}(t),\quad\quad Q(t)\in
\GR_{\Bk};
\ee
remember from (1.17), $Q(t)\in
\GR_{\Bk}$ means: $Q(t)$ is lower-triangular, with along the ``diagonal" 2$\times$2
matrices $c_{2n}I$, with $c_{2n}\neq 0$.

 In this section, we give the properties of the wave vectors and
their bilinear relations. Upon setting
\be
Q_1=Q(t)\mbox{\,\,and\,\,}Q_2=JQ^{\top -1}(t),
\ee
the matrix $Q(t)$ defines wave operators
\be
W_1(t)=Q_1(t)e^{\sum_1^{\iy}t_i\Lb^i},\quad W_2(t)=Q_2(t)e^{-\sum_1^{\iy}
t_i\Lb^{\top i}}=JW_1^{-1\top}(t),
\ee
$L$-matrices
\be
L:=L_1:=Q_1\Lb Q^{-1}_1,\quad L_2:=-\JR(L_1)=Q_2\Lb^{\top}Q_2^{-1},
\ee
and wave and dual wave vectors
\be
\begin{array}{lll}
\Psi_1(t,z)=W_1(t)\chi(z)&
&\Psi_1^{\ast}(t,z)=W_1^{-1}(t)^{\top}\chi(z^{-1})
 =-J\Psi_2(t,z^{-1})\\[4pt]
\Psi_2(t,z)=W_2(t)\chi(z)&
&\Psi_2^{\ast}(t,z)=W_2^{-1}(t)^{\top}\chi(z^{-1})=J\Psi_1(t,z^{-1}).
\end{array}
\ee
From the definition, it follows that the wave functions $\Psi_1$ have the
following asymptotics
$$
\left\{\begin{array}{ll}
\Psi_{1,2n}(t,z):=e^{\sum t_kz^k}z^{2n}
c_{2n}(t)\psi_{1,2n}(t,z),& \psi_{1,2n}=1+O(z^{-1})\\[4pt]
\Psi_{1,2n+1}(t,z)=e^{\sum
t_kz^k}z^{2n+1}c_{2n}(t)\psi_{1,2n+1}(t,z),&
\psi_{1,2n+1}=1+O(z^{-2})
\end{array}\right.
$$

\be
\left\{\begin{array}{ll}
\Psi_{2,2n}(t,z)=
e^{-\sum t_kz^{-k}}z^{2n+1}c^{-1}_{2n}(t)
\psi_{2,2n}(t,z),&
\psi_{2,2n}=1+O(z)\\[4pt]
\Psi_{2,2n+1}(t,z)=e^{-\sum
t_kz^{-k}}z^{2n}(-c_{2n}^{-1}(t))\psi_{2,2n+1}(t,z),&
\psi_{2,2n+1}=1+O(z^2),
\end{array}\right.
\ee
where the $c_i$ are the elements of the diagonal part of $Q$.

\begin{theorem} The $Q_i$, $L_i$ and $\Psi_i$ satisfy the equations
\be
\frac{\pl Q_1}{\pl t_i}=-(\pi_{\Bk}L_1^i)Q_1\quad\quad\frac{\pl Q_2}{\pl
t_i}=-(\JR(\pi_{\Bk}L_1^i))Q_2=(\pi_{\Bk_{+}}L_2^i)Q_2
\ee

\be
\frac{\pl L_1}{\pl t_i}=[-\pi_{\Bk}L_1^i,L_1]\quad\quad\frac{\pl L_2}{\pl t_i}
=[\pi_{\Bk_+}L_2^i,L_2]
\ee
\be
L_1\Psi_1=z\Psi_1\quad \quad L_2\Psi_2=z^{-1}\Psi_2,
\ee
\be
\frac{\pl\Psi_1}{\pl t_i}=(\pi_{\Bn}L_1^i)\Psi_1\quad\quad\frac{\pl\Psi_2}{\pl
t_i}=-(L_2^i-\pi_{\Bk_+}L_2^i)\Psi_2=-(\pi_{\Bn_+}L_2^i)\Psi_2,
\ee
 with
$\Psi_i$ satisfying the following bilinear identity for all
$n,m\in\BZ$,
\be
\oint_{\iy}\Psi_{1,n}(t,z)\Psi_{2,m}(t',z^{-1})\frac{dz}{2\pi
iz}+\oint_0\Psi_{2,n}(t,z)\Psi_{1,m}(t',z^{-1})\frac{dz}{2\pi iz}=0.
\ee
\end{theorem}

For later use, we shall also consider the ``monic" wave functions, with
the factors $c_{2n}(t)$ removed, i.e.,
\be
\hat \Psi_1(t,z):=Q_0^{-1}\Psi_1\mbox{\,\,and\,\,}\hat\Psi_2(t,z):=Q_0\Psi_2
\ee
and the matrix $\hat L_1$, normalized so as to have 1's above the
main diagonal, with $\hat Q:=Q_0^{-1}Q$,
\medbreak
\bea\hat L_1&=&Q_0^{-1}L_1Q_0=
(Q_0^{-1}Q)\Lb(Q_0^{-1}Q)^{-1}=\hat Q \Lb \hat Q^{-1}, \nonumber\\
\hat L_2&=&Q_0L_2Q_0^{-1} =-Q_0\JR(L_1)Q_0^{-1}=-\JR(\hat L_1)
\eea
Then, in terms of the elements $\hat q_{ij}$ of the matrix $\hat
Q:=Q_0^{-1}Q$, one easily computes by conjugation, that $\hat L_1$
has the following block structure:
\medbreak
$$\hat L_1=Q_0^{-1}L_1Q_0= (Q_0^{-1}Q)\Lb(Q_0^{-1}Q)^{-1}
=
 \pmatrix{&\vdots& \cr
 ...&\hat L_{00}&\hat L_{01}&0&0&\cr
&\hat L_{10}&\hat L_{11}&\hat L_{12}&0&\cr &*&\hat L_{21} &\hat
L_{22}&\hat L_{23}&\cr
 &*&*
&\hat L_{32}&\hat L_{33}&...\cr & & & &\vdots& &
\cr},
$$
with
$$
\hat L_{ii}:=\pmatrix{\hat q_{2i,2i-1}& 1 \cr
 \hat q_{2i+1,2i-1}-\hat q_{2i+2,2i} & -\hat q_{2i+2,2i+1}\cr},~~
 \hat L_{i,i+1}:=\pmatrix{0& 0 \cr
 1 & 0\cr}
$$
\be
\hat L_{i+1,i}:=\pmatrix{*&-\hat q_{2i+2,2i+1}^2-\hat q_{2i+3,2i+1}+\hat
q_{2i+2,2i}&  \cr
 * &*\cr}.
 \ee

\begin{theorem} $\hat L_i,~\hat Q,~\hat \Psi_1,~\hat \Psi_2$ satisfy the
 following equations:
\be
\frac{\pl\hat Q}{\pl t_n}=-\left((\hat L_1^n)_--Q_0^{-2}J((\hat L_1^n)_+)
^{\top}JQ_0^2\right)\hat Q,
\ee
and
\be
\hat L_1 \hat\Psi_1=z\hat\Psi_1\quad \quad \hat L_2 \hat\Psi_2
=z^{-1} \hat\Psi_2,
\ee
with
$$
\frac{\pl}{\pl t_n}\hat \Psi_1(t,z)=\left((\hat L_1^n)_++(\hat
L_1^n)_0+Q_0^{-2}\JR((\hat L_1^n)_+)Q_0^2\right)\hat \Psi_1(t,z),
$$
\bea
\frac{\pl}{\pl t_n}\hat \Psi_2(t,z)&=&\JR\left((\hat L_1^n)_++(\hat
L_1^n)_0+Q_0^{-2}\JR((\hat
L_1^n)_+)Q_0^2\right)\hat\Psi_2(t,z)\nonumber\\ &=&-\left((\hat
L^n_2)_-+(\hat L^n_2)_0+Q^2_0\JR((\hat
L_2^n)_-)Q_0^{-2}\right)\hat\Psi_2(t,z).\nonumber
\eea
\end{theorem}

The proof of Theorem 2.1 hinges on the following matrix version of
the bilinear identities:

\begin{lemma} The matrices $W_1(t)$ and $W_2(t)$,
defined in (2.3), satisfy
\be
W_1(t)W_1(t')^{-1}=W_2(t)W_2(t')^{-1}.
 \ee
\end{lemma}

\proof
The solution to the equation (2.0) is given by
$$
m_{\iy}(t)=e^{\sum t_k\Lb^k}m_{\iy}(0)e^{\sum t_k\Lb^{\top k}}.
$$
Therefore skew-Borel decomposing $m_{\iy}(t)$ and $m_{\iy}(0)$, we
find
\be
Q^{-1}(0)JQ^{\top
-1}(0)=e^{-\sum t_i\Lb^i}Q^{-1}(t)JQ^{\top -1}(t)e^{-\sum t_i\Lb^{\top
i}}
\ee
and so, from the definition of $W_1$ and $W_2$,
\bea
 W_1^{-1}(0)W_2(0)&=&Q^{-1}(0)JQ^{\top -1}(0)\nonumber\\
 &=&\left(Q(t)e^{\sum_1^{\iy}t_i\Lb^i}\right)^{-1}J\left(Q(t)e^{\sum
t_i\Lb^i}\right)^{\top -1}~\mbox{using (2.18)}\nonumber\\
 &=&W_1(t)^{-1}JW_1(t)^{\top
-1}\nonumber\\ &=&W_1(t)^{-1}W_2(t),
\eea
implying the independence in $t$ of the right hand
 side of (2.19).
Therefore, we have
$$
W_1(t)^{-1}W_2(t)=W_1(t')^{-1}W_2(t'),\quad\mbox{for
all\,\,}t,t'\in\BC^{\iy},
$$
and so
$$
W_1(t)W_1^{-1}(t')=W_2(t)W_2^{-1}(t').
$$
\qed

\underline{\sl Proof of Theorem 2.1 }: The proof of equation
(2.7) for $Q_1$, namely
$$\frac{\pl Q_1}{\pl t_i}=-(\pi_{\Bk}L_1^i)Q_1,
$$ follows at once from Proposition 1.2.

The proof of (2.7) for $Q_2=JQ_1^{\top -1}$ is based on the
identity $\JR\pi_{\Bk}a=\pi_{\Bk_+}\JR a$. Indeed, we compute
\bea
\frac{\pl Q_2}{\pl t_i}Q_2^{-1}&=&-JQ_1^{\top -1}\frac{\pl Q_1^{\top}}{\pl
t_i}Q_1^{\top -1}Q_2^{-1}\nonumber\\ &=&-JQ_1^{\top -1}Q_1^{\top
}(\pi_{\Bk}L_1^i)^{\top}Q_1^{\top -1}Q_1^{\top }J\nonumber\\
&=&-J(\pi_{\Bk}L_1^i)^{\top}J\nonumber\\
&=&-\JR(\pi_{\Bk}L_1^i)\nonumber\\ &=&-\pi_{\Bk_+}\JR
L_1^i\nonumber\\ &=&-\pi_{\Bk_+}\JR(-\JR L_2)^i, ~~\mbox{using
(2.4)},\nonumber\\ &=&-\pi_{\Bk_+}\JR(-1)^i(\JR L_2)^i\nonumber\\
&=&-\pi_{\Bk_+}\JR(-1)^i(-1)^{i-1}\JR L_2^i\nonumber\\
&=&\pi_{\Bk_+}L_2^i.\nonumber
\eea
Equations (2.8) and (2.10) for $L_1,~L_2$ and $\Psi_1,~\Psi_2$ are
then straightforward.

Finally, the proof of the bilinear identity (2.11) proceeds as
follows: By a well-known lemma (see \cite{DJKM}) ,
$$W_{{1}\atop{2}}W_{{1}\atop{2}}
(t)W_{{1}\atop{2}}(t')^{-1}=
\oint_{{\iy}\atop{0}}\Psi_{{1}\atop{2}}(t,z)\otimes
\Psi^*_{{1}\atop{2}}(t',z)\frac{dz}{2\pi iz}
$$
and so the statement of Lemma 2.3 yields
$$
\oint_{\iy}\Psi_1(t,z)\otimes\Psi_1^*(t',z)\frac{dz}{2\pi
iz}=\oint\Psi_2(t,z)\otimes\Psi_2^*(t',z)\frac{dz}{2\pi iz},
$$
whose $(m,n)$th component is
$$
\oint_{\iy}\Psi_{1,n}(t,z)\Psi^*_{1,m}(t',z)
\frac{dz}{2\pi
iz}-\oint_0\Psi_{2,n}(t,z)\Psi^*_{2,m}(t',z)
\frac{dz}{2\pi iz}=0.
$$
 Next we use the relations
$\Psi^*_1(t,z)=-J\Psi_2(t,z^{-1})$
and $\Psi^*_2(t,z)=J\Psi_1(t,z^{-1})$, to yield
$$
\oint_{\iy}\Psi_1(t,z)\otimes J\Psi_2(t',z^{-1})\frac{dz}{2\pi
iz}+\oint_0\Psi_2(t,z)\otimes J\Psi_1(t',z^{-1})\frac{dz}{2\pi iz}=0,
$$
which again componentwise leads to (2.11).\qed

\vspace{0.5cm}

\underline{\sl Proof of Theorem 2.2 }: To prove (2.15), remember from Theorem 2.1,
$$
\frac{\pl Q}{\pl t_n}Q^{-1}=-\pi_{\Bk}L^n=-((L^n)_-+J(L^n_+)^{\top}J)-\frac{1}{2}
((L^n)_0-J((L^n)_0)^{\top}J);
$$
hence, taking the $(\,)_0$-part of this expression, yields
$$
\frac{\pl \log Q_0}{\pl t_n}=\left(\frac{\pl Q}{\pl
t_n}Q^{-1}\right)_0=-\pi_{\Bk}(L^n)_0=-\frac{1}{2}
(L^n)_0+\frac{1}{2}J(L^n)_0^{\top}J.
$$
Using the fact that $Q_0,Q_0^{-1},\dot Q_0\in G_k\cap\DR_0$ commute
among themselves and commute with $J$ and the fact that
$\DR_0\DR_{\pm}$, $\DR_{\pm}\DR_0\subset\DR_{\pm}$, we compute for
$\hat Q=Q_0^{-1} Q,~ \hat L_1= Q_0^{-1} L_1 Q_0,$ (see (2.13))
\bea
\frac{\pl \hat Q}{\pl t_n}\hat Q^{-1}&=&-Q_0^{-1}\dot Q_0Q_0^{-1}QQ^{-1}Q_0+Q_0^{-1}\dot QQ^{-1}Q_0\nonumber\\
&=&-Q_0^{-1}\dot Q_0+Q_0^{-1}\dot QQ^{-1}Q_0\nonumber\\
&=&Q_0^{-1}\left(-\dot Q_0Q_0^{-1}+\dot
QQ^{-1}\right)Q_0\nonumber\\
&=&Q_0^{-1}\left(-(L_1^n)_-+J(L^n_{1+})^{\top}J\right)Q_0\nonumber\\
&=&-(Q_0^{-1}L_1^nQ_0)_-+Q_0^{-1}J
\left(Q_0(Q_0^{-1}L_1^nQ_0)_+Q_0^{-1}\right)^{\top}
JQ_0\nonumber\\ &=&-(\hat L_1^n)_-+Q_0^{-2}J((\hat
L_1^n)_+)^{\top}JQ_0^2.\nonumber
\eea
Using this result and $\hat L_1\hat \Psi_1(t,z)=z \hat
\Psi_1(t,z)$, we find
\bea
\lefteqn{
\frac{\pl \hat \Psi_1(t,z)}{\pl t_n}
}\nonumber\\
&=&\frac{\pl}{\pl t_n}e^{\sum t_iz^i}\hat Q\chi(z)\nonumber\\
&=&z^ne^{\sum_1^{\iy}t_iz^i}\hat
Q\chi(z)+e^{\sum_1^{\iy}t_iz^i}\left(-(\hat L_1^n)_-
+Q_0^{-2}J((\hat L_1^n)_+)^{\top}JQ_0^2\right)\hat
Q\chi(z)\nonumber\\ &=&(\hat L_1^n-(\hat L_1^n)_-+Q_0^{-2}J((\hat
L_1^n)_+)^{\top}JQ_0^2)\hat\Psi_1(t,z)\nonumber\\ &=&((\hat
L_1^n)_++(\hat L_1^n)_0+Q_0^{-2}(\JR(\hat
L_1^n)_+)Q^2_0)\hat\Psi_1(t,z).
\eea
But, we also have that $\hat \Psi_1=Q_0^{-1}\Psi_1(t,z)$ and $\hat
\Psi_2=Q_0\Psi_2(t,z)$ satisfy, using $W_2=JW_1^{-1 \top}$,
\bea
\frac{\pl \hat\Psi_1(t,z)}{\pl t_n}=(Q_0^{-1}W_1)^{.}\chi(z)&=&
 (Q_0^{-1}W_1)^{.} (Q_0^{-1}W_1)^{-1} \hat\Psi_1(t,z)\\
\frac{\pl \hat\Psi_2(t,z)}{\pl t_n}=(Q_0W_2)^{.}\chi(z)&=&(Q_0W_2)
^{.}(Q_0W_2)^{-1}
(Q_0\Psi_2)\nonumber\\ &=&(\dot Q_0W_2+Q_0\dot
W_2)W_2^{-1}Q^{-1}_0(Q_0\Psi_2)\nonumber\\ &=&(\dot
Q_0Q^{-1}_0+Q_0\dot W_2W_2^{-1}Q^{-1}_0)Q_0\Psi_2\nonumber\\
&=&(\dot Q_0Q_0^{-1}+Q_0\JR(\dot
W_1W_1^{-1})Q_0^{-1})Q_0\Psi_2\nonumber\\ &=&(\dot
Q_0Q_0^{-1}+Q_0J(\dot
W_1W_1^{-1})^{\top}JQ_0^{-1})Q_0\Psi_2\nonumber\\ &=&(-J\dot
Q_0Q_0^{-1}+J(Q^{-1}_0\dot
W_1W_1^{-1}Q_0)^{\top})JQ_0\Psi_2\nonumber\\ &=&J(-Q_0^{-1}\dot
Q_0+Q_0^{-1}\dot W_1W_1^{-1}Q_0)^{\top}J Q_0\Psi_2\nonumber\\
&=&\JR\left((Q^{-1}_0W_1)^{.}(Q_0^{-1}W_1)^{-1}\right)(Q_0\Psi_2).
 \eea
Comparing (2.21), (2.22) and (2.23), and using
$$
-\JR(\hat L_1^n)=\hat L_2^n,
$$
and so, in particular,
$$
-\JR((\hat L_{1 }^n)_{\pm})=(\hat L_{2 }^n)_{\mp}~~\mbox{and}~~
-\JR((\hat L_{1}^n)_{0})=(\hat L_{2 }^n)_0,
$$
$$
\frac{\pl \hat\Psi_2(t,z)}{\pl t_n}=-((\hat L_2^n)_-+(\hat
L_2^n)_0+Q^2_0(\JR(L^n_2)_-)Q_0^{-2})\hat\Psi_2(t,z),
$$ which establishes theorem 2.2.
\qed

\section{Existence of the Pfaff $\tau$-function}

The point of this section is to show that the solution of the Pfaff
Lattice can be expressed in terms of a sequence of functions
$\tau$, which are not $\tau$-functions in the usual sense, but
enjoys a different set of bilinear identities and partial
differential equations.

\begin{proposition}There exists functions $\tau_{2n}(t)$
 such that
\be
 \psi_{1,2n}(t,z)=\frac{\tau_{2n}(t-[z^{-1}])}{\tau_{2n}(t)}~
 \mbox{ and }~
\psi_{2,2n}(t,z)=\frac{\tau_{2n+2}(t+[z])}{\tau_{2n+2}(t)}.
\ee
\end{proposition}

\bigbreak

The proof of proposition 3.1 will be postponed until later. For
future use, we define the diagonal matrix
\be   h=\mbox{diag}(...,h_{-2},h_{-2}, h_0, h_0, h_2,
 h_2,...)\in \DR_0^-,~~\mbox{with}~~ h_{2n}=
 \frac{ \tau_{2n+2}}{ \tau_{2n}}.
\ee

\begin{theorem}

\bea
\Psi_{1,2n}(t,z)&=&e^{\sum t_i z^i}z^{2n}
\frac{\tau_{2n}(t-[z^{-1}])}
{\sqrt{\tau_{2n}(t)\tau_{2n+2}(t)}}
\nonumber\\
 \Psi_{1,2n+1}(t,z)&=&e^{\sum t_i z^i}z^{2n}
\frac{(z+\pl/ \pl t_1)\tau_{2n}(t-[z^{-1}])}
{\sqrt{\tau_{2n}(t)\tau_{2n+2}(t)}}
\nonumber\\
 \Psi_{2,2n}(t,z)&=&e^{-\sum t_i z^{-i}}z^{2n+1}
\frac{\tau_{2n+2}(t+[z])}
{\sqrt{\tau_{2n}(t)\tau_{2n+2}(t)}}
\nonumber\\
 \Psi_{2,2n+1}(t,z)&=&e^{-\sum t_i z^{-i}}z^{2n+1}
\frac{(z^{-1}-\pl/ \pl t_1)\tau_{2n+2}(t+[z])}{\sqrt{\tau_{2n}(t)\tau_{2n+2}(t)}}
,\nonumber\\
 \eea
with the $\tau_{2n}(t)$ satisfying the following bilinear identity
\bea
& &\oint_{z=\iy}\tau_{2n}(t-[z^{-1}])\tau_{2m+2}(t'+[z^{-1}])
e^{\sum(t_i-t'_i)z^i} z^{2n-2m-2}\frac{dz}{2\pi i}\nonumber\\
&&\quad+\oint_{z=0}\tau_{2n+2}(t+[z])\tau_{2m}(t'-[z])
e^{\sum(t'_i-t_i)z^{-i}}z^{2n-2m}\frac{dz}{2\pi i}=0.
\eea
Then $L$ has the following representation in terms of the Pfaffian
$\tau$-functions:
\bigbreak
$$ h^{1/2} L  h^{-1/2}=
 \pmatrix{&\vdots& \cr
 ...&\hat L_{00}&\hat L_{01}&0&0&\cr
&\hat L_{10}&\hat L_{11}&\hat L_{12}&0&\cr &*&\hat L_{21} &\hat
L_{22}&\hat L_{23}&\cr
 &*&*
&\hat L_{32}&\hat L_{33}&...\cr & & & &\vdots& &
\cr},
$$
with $({}^.=\frac{\pl}{\pl t_1})$
$$
\hat L_{nn}:=\pmatrix{-(\log \tau_{2n})^. & &1 \cr \cr
 -\frac{S_2(\tilde \pl)\tau_{2n}}{\tau_{2n}}
-\frac{S_2(-\tilde \pl)\tau_{2n+2}}{\tau_{2n+2}} & &
 (\log \tau_{2n+2})^.\cr}~~
 ~~~~~~\hat L_{n,n+1}:=\pmatrix{0& 0 \cr
 1 & 0\cr}
$$

\vspace{0.5cm}

\be
\hat L_{n+1,n}:=\pmatrix{*&(\log \tau_{2n+2})^{..}&  \cr
 * &*\cr}.
 \ee

\end{theorem}

The following bilinear relations are due to \cite{ASV}:

\begin{corollary}
The functions $\tau_{2n}(t)$ satisfy the following ``differential
Fay identity"\footnote{$\{f,g\}=f'g-fg'$, where $'=\pl/\pl t_1$.}

\bigbreak

$
\{\tau_{2n}(t-[u]),\tau_{2n}(t-[v])\}
$\hfill
\bea
& & \hspace{2cm}+(u^{-1}-v^{-1})(\tau_{2n}(t-[u])\tau_{2n}(t-[v])
-\tau_{2n}(t)\tau_{2n}(t-[u]-[v]))\nonumber\\
& & \nonumber\\
 & &~~~=uv(u-v)\tau_{2n-2}(t-[u]-[v])\tau_{2n+2}(t),
\eea
and Hirota type bilinear equations, always involving nearest
neighbours:
\be
\left(p_{k+4}(\tilde\pl)-\frac{1}{2}\frac{\pl^2}{\pl
t_1\pl t_{k+3}}\right)\tau_{2n}\circ\tau_{2n}=p_k(\tilde
\pl)~\tau_{2n+2}\circ\tau_{2n-2}
\ee
\hfill $k,n=0,1,2,...~.$
\end{corollary}

\begin{lemma} Consider an arbitrary function $\vp(t,z)$ depending on
$t\in\BC^{\iy}$, $z\in\BC$, having the asymptotics
$\vp(t,z)=1+O(\frac{1}{z})$ for $z\nearrow\iy$ and
 satisfying the functional
relation
$$
\frac{\vp(t-[z^{-1}_2],z_1)}{\vp(t,z_1)}=\frac{\vp(t-[z^{-1}_1],z_2)}
{\vp(t,z_2)}, \quad t\in\BC^{\iy},z\in\BC.
$$
Then there exists a function $\tau(t)$ such that
$$\vp(t,z)=\frac{\tau(t-[z^{-1}])}{\tau(t)}.
$$
\end{lemma}

\proof See appendix (Section 10)

\begin{lemma} The following holds for the
Pfaffian wave function $\Psi_1$ and $\Psi_2$, as in (2.6),
\be
\frac{\psi_{1,2n}(t-[z^{-1}_2],z_1)}{\psi_{1,2n}(t,z_1)}=
\frac{\psi_{1,2n}(t-[z^{-1}_1],z_2)}{\psi_{1,2n}(t,z_2)}
\ee
and
\be
\psi_{2,2n-2}(t-[z^{-1}],z^{-1})\psi_{1,2n}(t,z)=1.
\ee
\end{lemma}

\proof Setting (2.6) in the bilinear equation (2.11), with $n\mapsto 2n,~m\mapsto 2n-2$, yields
$$
\frac{c_{2n}(t)}{c_{2n-2}(t)}\oint_{\iy}e^{\sum
(t_i-t'_i)z^i}\psi_{1,2n}(t,z)\psi_{2,2n-2}(t',z^{-1})\frac{dz}{2\pi i}
$$
$$
+\frac{c_{2n-2}(t)}{c_{2n}(t)}\oint_0 e^{\sum
(t'_i-t_i)z^i}\psi_{2,2n}(t,z)\psi_{1,2n-2}(t',z^{-1})\frac{z^2dz}{2\pi
i}=0.
$$
Setting
$$
t-t'=[z^{-1}_1]+[z^{-1}_2]
$$
in the above and using $e^{\sum_1^{\iy}x^i/i}=1/(1-x)$ yields
$$
\frac{c_{2n}}{c_{2n-2}}\oint_{\iy}\frac{\psi_{1,2n}(t,z)\psi_{2,2n-2}(t',z^{-1})}
{\left(1-\frac{z}{z_1}\right)\left(1-\frac{z}{z_2}\right)}
\frac{dz}{2\pi i}\hspace{6cm}
$$
$$
\hspace{1cm}=-\frac{c_{2n-2}}
{c_{2n}}\oint_{z=0}z^2\left(1-\frac{z}{z_1}\right)\left(1-\frac{z}{z_2}\right)
\psi_{2,2n}(t,z)\psi_{1,2n-2}(t',z^{-1})\frac{dz}{2\pi i}.$$

\medbreak
\noindent Since the integrand on the right hand side is holomorphic, it suffices to
evaluate the integral on the left hand side, which can be viewed as an
integral along a contour encompassing $\iy$ and the points $z_1$ and $z_2$,
thus leading to
\be
\psi_{1,2n}(t,z_1)\psi_{2,2n-2}(t-[z_1^{-1}]-[z_2^{-1}],z_1^{-1})=\psi_{1,2n}(t,z_2)\psi_{2,2n-2}(t-
[z_1^{-1}]-[z_2^{-1}],z_2^{-1})
\ee
with
$$
\psi_{1,2n}(t,z)=1+O\left(z^{-1}\right),~~~
\psi_{2,2n-2}(t-[z_1^{-1}]-[z_2^{-1}],z^{-1})
=1+O(z^{-2}).
$$
Therefore, letting $z_2\nearrow\iy$, one finds
\be
\psi_{1,2n}(t,z_1)\psi_{2,2n-2}(t-[z_1^{-1}],z_1^{-1})=1
,\ee
 yielding (3.9), and so, upon shifting $t\mapsto t-[z_2^{-1}]$,
$$
\psi_{2,2n-2}(t-[z_1^{-1}]-[z_2^{-1}],z_1^{-1})=\frac{1}{\psi_{1,2n}(t-[z_2^{-1}],z_1)};
$$
similarly
\be
\psi_{2,2n-2}(t-[z_1^{-1}]-[z_2^{-1}],z_2^{-1})=\frac{1}{\psi_{1,2n}(t-[z_1^{-1}],z_2)}.
\ee
Setting the two expressions (3.12) in (3.10) yields
$$
\frac{\psi_{1,2n}(t-[z_2^{-1}],z_1)}{\psi_{1,2n}(t,z_1)}=\frac{\psi_{1,2n}(t-[z_1^{-1}],z_2)}
{\psi_{1,2n}(t,z_2)}.
$$
\qed

\underline{\sl Proof of Proposition 3.1 }: From Lemmas 3.4 and 3.5, there exists,
for each $2n$, a function $\tau_{2n}$ such that the first relation of (3.1)
is satisfied, i.e.,
$$
\psi_{1,2n}(t,z)=\frac{\tau_{2n}(t-[z^{-1}])}{\tau_{2n}(t)},
$$
and so from (3.9)
$$
\psi_{2,2n-2}(t-[z^{-1}],z^{-1})=\frac{1}{\psi_{1,2n}(t,z)}=\frac{\tau_{2n}(t)}{\tau_{2n}(t-[z^{-1}])},
$$
thus leading to
$$
\psi_{2,2n-2}(t,z)=\frac{\tau_{2n}(t+[z])}{\tau_{2n}(t)},
$$
which is the second relation of (3.1).\qed

\underline{\sl Proof of Theorem 3.2 }:  At first, remembering that
$\hat Q=Q_0^{-1}Q$, observe that
\bea
e^{\sum t_iz^i}((\hat
Q)\chi(z))_{2n}&=&(Q^{-1}_0\Psi_1(t,z))_{2n}\nonumber\\ &=&e^{\sum
t_iz^i}z^{2n}\psi_{1,2n}(t,z)\nonumber\\ &=&e^{\sum
t_iz^i}z^{2n}\frac{\tau_{2n}(t-[z^{-1}])}{\tau_{2n}(t)}\nonumber\\
&=&e^{\sum
t_iz^i}z^{2n}\left(1+\sum_{n=1}^{\iy}\frac{S_k(-\tilde\pl)\tau_{2n}(t)}{\tau_{2n}(t)}\right),\nonumber
\eea
showing that a few subdiagonals of the matrix $\hat Q$ are
 given by
$$
\hat Q=\left(\begin{array}{cc@{}c@{}ccc}
 \ddots& &&&&\\
  &\boxed{\begin{array}{cc}~~~  1 ~~~~~~&~~~~~~~~ 0~~~~ \\
~~~ 0~~~~~~&~~~~~~~~1~~~~ \end{array}}& & & &\\
 &\boxed{\begin{array}{cc}  \hat q_{2n,2n-2}&\hat q_{2n,2n-1} \\
 \hat q_{2n+1,2n-2}&\hat q_{2n+1,2n-1} \end{array}}&
  \boxed{\begin{array}{cc} ~~~~  1 ~~~~~~&~~~~~~~~ 0~~~~ \\
~~~~ 0~~~~~~&~~~~~~~~1~~~~ \end{array}} &&&\\
 & &&& \ddots
 \end{array}
 \right)
 $$
with
\be
\hat q_{2n,2n-1}=-\frac{\pl}{\pl t_1}\log\tau_{2n},\quad\hat
q_{2n,2n-2}=\frac{S_2(-\tilde\pl)\tau_{2n}}{\tau_{2n}}.
\ee
Remembering that
\be
\left\{\begin{tabular}{ll}
$\hat\Psi_{1,2n}(t,z)=e^{\sum t_kz^k}z^{2n}
\psi_{1,2n}(t,z),$& $\psi_{1,2n}=1+O(z^{-1})$\\
$\hat \Psi_{1,2n+1}(t,z)=e^{\sum
t_kz^k}z^{2n+1}\psi_{1,2n+1}(t,z),$& $\psi_{1,2n+1}=1+O(z^{-2})$
\end{tabular}\right.
\ee
\bea
\left\{\begin{tabular}{ll}
$\hat \Psi_{2,2n}(t,z)= e^{-\sum t_kz^{-k}}z^{2n+1}
\psi_{2,2n}(t,z),$&
$\psi_{2,2n}=1+O(z)$\nonumber\\
 $\hat \Psi_{2,2n+1}(t,z)=-e^{-\sum
t_kz^{-k}}z^{2n}\psi_{2,2n+1}(t,z),
 $& $\psi_{2,2n+1}=1+O(z^2),$
\end{tabular}\right.\nonumber\\
\eea
we now compute, using theorem 2.2,
\bea
\lefteqn{
e^{\sum t_iz^i}\left(\frac{\pl}{\pl
t_1}+z\right)z^{2n}\psi_{1,2n}(t,z)
}
\nonumber\\
&=&\left(\frac{\pl}{\pl t_1}\hat
\Psi_1(t,z)\right)_{2n}\nonumber\\
&=&\left(\left((\hat
L_1)_++(\hat L_1)_0+Q_0^{-2}J(\hat
L_{1+})^{\top}JQ_0^2\right)\hat\Psi_1(t,z)\right)_{2n}
\eea
and
\bea
-\lefteqn{
e^{-\sum t_iz^{-i}}\left(\frac{\pl}{\pl
t_1}-\frac{1}{z}\right)z^{2n+1}\psi_{2,2n}(t,z)
}
\nonumber\\
&=&\frac{\pl}{\pl t_1}(\hat\Psi_2(t,z))_{2n}\nonumber\\
&=&\left(\left(\JR((\hat L_1)_++(\hat L_1)_0 +Q_0^{-2}J(\hat
L_{1+})^{\top}JQ_0^2)\right)\hat \Psi_2(t,z)\right)_{2n}.
\eea
In this expression, the matrix equals, according to (2.13),

\medbreak
\bean
\lefteqn{(\hat L_1)_++(\hat L_1)_0+Q_0^{-2}J(\hat
L_{1+})^{\top}JQ_0^2=}
\\
&&
\pmatrix{&\vdots& \cr
 ...&\hat q_{0,-1}&1&\vline&0&0&\vline&0&0\cr
&\hat q_{1,-1}-\hat q_{20}&-\hat
q_{21}&\vline&1&0&\vline&0&0\cr\hline
 &0 & 0
&\vline&\hat q_{21}&1&\vline&0&0\cr &{c_0^2}/{c_2^2}&0&\vline&\hat
q_{31}-\hat q_{42}&-\hat q_{43}&\vline&1&0\cr\hline
 &0&0&\vline&0&0&\vline&\hat q_{43}&1\cr
  &0&0&\vline&{c_2^2}/{c_4^2} &0&\vline&\hat q_{53}-\hat
q_{64}&-\hat q_{65}&...\cr
 & & & & & & & &\vdots
\cr},
\eean
and, acting with $\JR$ on this matrix,

\medbreak
\noindent$\displaystyle{\JR\left((\hat L_1)_++(\hat L_1)_0+Q_0^{-2}J(\hat
L_{1+})^{\top}JQ_0^2\right)=}$

$$
\pmatrix{&\vdots& \cr
 ...&-\hat q_{21}&1&\vline&0&0&\vline&0&0\cr
&\hat q_{1,-1}-\hat q_{20}&\hat
       q_{0,-1}&\vline&c^2_0/c^2_2&0&\vline&0&0\cr\hline
 &0 & 0&\vline&-\hat q_{43}&1&\vline&0&0\cr
 &1&0&\vline&\hat q_{31}-\hat
q_{42}&\hat q_{21}&\vline&c_2^2/c^2_4&0\cr\hline
  &0&0&\vline&0&0&\vline&-\hat
q_{65}&1\cr
 &0 &0&\vline&1&0&\vline&\hat q_{53}-\hat q_{64}&\hat
q_{43}&...\cr & & & & & & & & \vdots \cr},
$$
using the fact that
$$
\JR\pmatrix{0&0\cr
1&0\cr}=\pmatrix{0&0\cr
1&0\cr}.
$$
Therefore the $2n$th rows of both matrices respectively have the
form
$$(0,\dots,0,\begin{array}[t]{@{}c@{}}
\hat q_{2n,2n-1}(t)\\\stackrel{\uparrow}{2n}\end{array},1,0,0,\dots)$$
$$(0,\dots,0,\begin{array}[t]{@{}c@{}}
-\hat q_{2n+2,2n+1}(t)\\\stackrel{\uparrow}{2n}\end{array},1,0,0,\dots)$$
and thus from (3.16) and (3.17), and the expansions (3.14) and
(3.15), we have
\bea
\left(\frac{\pl}{\pl t_1}+z\right)\psi_{1,2n}(t,z)&=&\hat q_{2n,2n-1}(t)\psi_{1,2n}+z\psi_{1,2n+1}
\nonumber\\
\left(\frac{\pl}{\pl t_1}-z^{-1}\right)\psi_{2,2n}(t,z)&=&\hat q_{2n+2,2n+1}(t)
\psi_{2,2n}+z^{-1}\psi_{2,2n+1}
\eea
and so, using the expression (3.13) for $\hat q_{2n,2n-1}$ and the
first expression (3.1),

\medbreak

\noindent
$\displaystyle{z^{2n+1}\psi_{1,2n+1}(t,z)}$
\bea
&=&\left(z+\frac{\pl}{\pl t_1}\right)z^{2n}\psi_{1,2n}(t,z)-\hat
q_{2n,2n-1}(t)z^{2n}\psi_{1,2n}(t,z)\nonumber\\
&=&\left(z+\frac{\pl}{\pl
t_1}\right)z^{2n}\psi_{1,2n}(t,z)+\left(\frac{\pl}{\pl
t_1}\log\tau_{2n}(t)\right)z^{2n}\psi_{1,2n}(t,z)\nonumber\\
&=&\left(z+\frac{\pl}{\pl
t_1}\right)z^{2n}\frac{\tau_{2n}(t-[z^{-1}])}{\tau_{2n}(t)}+\left(\frac{\pl}{\pl
t_1}\tau_{2n}(t)\right)z^{2n}\frac{\tau_{2n}(t-[z^{-1}])}{\tau^2_{2n}(t)}\nonumber\\
&=&z^{2n}\frac{\left(z+\frac{\pl}{\pl
t_1}\right)\tau_{2n}(t-[z^{-1}])}{\tau_{2n}(t)}
\eea
and similarly, using the relation (3.17),
\be
z^{2n}\psi_{2,2n+1}(t,z)=z^{2n+1}\frac{\left(-z^{-1}+\frac{\pl}{\pl
t_1}\right)\tau_{2n+2}(t+[z])}{\tau_{2n+2}(t)}.
\ee
Therefore, we also have
\bea
\psi_{1,2n+1}(t,z)&=&\frac{1}{z}\frac{1}{\tau_{2n}(t)}\left(z+\frac{\pl}{\pl
t_1}\right)\left(\tau_{2n}(t)-\frac{\pl\tau_{2n}}{\pl
t_1}z^{-1}+S_2(-\tilde\pl)\tau_{2n}z^{-2}+\cdots\right)\nonumber\\
&=&1+\frac{1}{\tau_{2n}(t)}\left(-\frac{\pl^2}{\pl
t_1^2}+S_2(-\tilde\pl)\right)\tau_{2n}z^{-2}+O(z^{-3})\nonumber
\eea
thus
\be
\hat q_{2n+1,2n}=0,\quad
\hat q_{2n+1,2n-1}=\frac{1}{\tau_{2n}}\left(S_2(-\tilde\pl)-\frac{\pl^2}
{\pl
t_1^2}\right)\tau_{2n}=\frac{-S_2(\tilde\pl)\tau_{2n}}{\tau_{2n}}.
\ee

Setting $n\mapsto 2n$ and $m\mapsto 2n$ in the bilinear relation
(2.11) and substituting, using (2.6) and the expressions for
$\psi_{1,2n}(t,z)$ and $\psi_{2,2n}(t,z)$ in the proof of
proposition 3.1,
$$
\Psi_{1,2n}(t,z)=e^{\sum
t_kz^k}z^{2n}c_{2n}(t)\frac{\tau_{2n}(t-[z^{-1}])}{\tau_{2n}(t)}
$$
and
$$
\Psi_{2,2n}(t',z)=e^{-\sum
t_kz^{-k}}z^{2n+1}c_{2n}^{-1}(t')\frac{\tau_{2n+2}(t+[z])}{\tau_{2n+2}(t)}
$$
into
$$
\oint_{\iy}\Psi_{1,2n}(t,z)\Psi_{2,2n}(t',z^{-1})\frac{dz}{2\pi iz}+
\oint_0\Psi_{2,2n}(t,z)\Psi_{1,2n}(t',z^{-1})\frac{dz}{2\pi iz}=0
$$
yields
$$
\frac{c_{2n}(t)}{c_{2n}(t')}\oint_{\iy}e^{\sum(t_k-t'_k)z^k}
\frac{\tau_{2n}(t-[z^{-1}])\tau_{2n+2}(t'+[z^{-1}])}{\tau_{2n}(t)
\tau_{2n+2}(t')}\frac{dz}{2\pi iz^2}
$$
$$
+\frac{c_{2n}(t')}{c_{2n}(t)}\oint_0e^{\sum(t'_k-t_k)z^{-k}}
\frac{\tau_{2n+2}(t+[z])\tau_{2n}(t'-[z])}{\tau_{2n+2}(t)
\tau_{2n}(t')}\frac{dz}{2\pi i}.
$$
Setting $t'=t+[\al]$ amounts to replacing the exponential:
$$
e^{\sum(t_k-t'_k)z^k}=1-\al z,~\quad
e^{\sum(t'_k-t_k)z^{-k}}=\frac{1}{1-\al/z},
$$
so that the first integral has a simple pole at $z=\iy$ and the second
integral one at $z=\al$. Evaluating the integrals yield
$$
-\al\frac{c^2_{2n}(t)\frac{\tau_{2n+2}(t)}
 {\tau_{2n}(t)}}{c^2_{2n}(t')
\frac{\tau_{2n+2}(t')}{\tau_{2n}(t')}}+\al=0;
$$
i.e.,
$$
\left(e^{\sum\frac{\al^i}{i}\frac{\pl}{\pl t_i}}-1\right)
c_n^2(t)\frac{\tau_{2n+2}(t)}{\tau_{2n}(t)}=0
$$
yielding the following relation, which involves a constant $c_n$,
independent of time,
\be
c_{2n}^2(t)=c_n\frac{\tau_{2n}(t)}{\tau_{2n+2}(t)}=c_n\cdot
h^{-1}_{2n}(t).
\ee
Rescaling $\tau_{2n}\mapsto \tau_{2n}/(c_1 c_2\cdots c_{n-1}) $
yields (3.3). Using the expressions for $\psi_{1,2n}(t,z)$ and
$\psi_{2,2n}(t,z)$ (see the proof of proposition
 3.1), (3.19), (3.20), (3.21), (2.6) and substituting
 (3.3) into (2.11) yields (3.4).

Finally to derive the form (3.5) of the matrix $L$, set (3.13) and
(3.20) in the elements just below the main diagonal of matrix
(2.14), to yield ($^.=\pl/\pl t_1$)
\bea
\lefteqn{- \hat q_{2n,2n-1}^2- \hat q_{2n+1,2n-1}+
\hat q_{2n,2n-2}}
\nonumber\\
&=&-\left(\frac{\dot{
 \tau}_{2n}}{ \tau_{2n}}\right)^2-
 \frac{(S_2(-\tilde\pl)-
\frac{\pl^2}{\pl t_1^2})\tau_{2n}}{\tau_{2n}}+
\frac{S_2(-\tilde\pl)\tau_{2n}}{\tau_{2n}}\nonumber\\
&=&\frac{\ddot\tau_{2n}}{\tau_{2n}}-\left(\frac{\dot\tau_{2n}}{\tau_{2n}}\right)^2\nonumber\\
&=&(\log\tau_{2n})^{..}\nonumber
\eea
and
\bea
 \hat q_{2n+1,2n-1}-\hat q_{2n+2,2n}&=&\frac{(S_2(-\tilde\pl)-
\frac{\pl^2}{\pl
t_1^2})\tau_{2n}}{\tau_{2n}}-\frac{S_2(-\tilde\pl)\tau_{2n+2}}{\tau_{2n+2}}\nonumber\\
&=&-\frac{S_2(\tilde\pl)\tau_{2n}}{\tau_{2n}}-\frac{S_2(-\tilde\pl)
\tau_{2n+2}}{\tau_{2n+2}},\nonumber
\eea
concluding the proof of theorem 3.2, upon substituting these
relations into (2.14).\qed


\section{Semi-infinite matrices $m_{\iy}$,
(skew-)orthogonal polynomials and matrix integrals}

\subsection{$\pl m / \pl t_k=\Lb^k m $, orthogonal polynomials and
Hermitean matrix integrals. }

 For the sake of completeness and analogy, we add this
  subsection, which summarizes some of \cite{AvM2}. Consider a $t$-dependent weight
$\rho_t(dz):=e^{-V_t(z)}dz:=e^{-V(z)+\sum t_i z^i}dz=e^{\sum t_i
z^i}\rho(dz)$ on $\BR$, as in (0.0) and the induced $t$-dependent
measure
\be
e^{Tr (- V(X)+\sum t_i X^i)}dX,
\ee
on the ensemble
$
{\cal H}_n$ of Hermitean matrices, with Haar measure $dX$; the
latter can be decomposed into a spectral part (radial part) and an
angular part:
\be
dX:=\displaystyle{\prod_1^n dX_{ii}\prod_{1\leq i<j\leq n}}(d\Re
X_{ij}\,d\Im X_{ij})=\Delta^2(z)dz_1\cdots dz_n~dU,
\ee
where $\Delta(z)=\displaystyle{\prod_{1\leq i<j\leq n}}(z_i-z_j)$
is the Vandermonde determinant. Here we form the following matrix
integral
\be
\int_{\HR_n}e^{Tr (- V(X)+\sum t_i X^i)}dX=c_n
\int_{\BR^n}\Delta^2(z)\prod^n_1\rho_t(dz_k).
\ee
The weight $\rho_t(dz)$ defines a (symmetric) inner product
$$
\la f,g\ra^{sy}=\int f(z)g(z)\rho_t(dz)
$$
and so, the moments
$$
\mu_{ij}(t):=\la
z^i,z^j\ra^{sy}=
\int_{\BR} z^{k+\ell}e^{\sum t_iz^i}\rho(dz)=
\mu_{i+\ell,j}(t)
$$
satisfy
$$
\frac{\pl\mu_{ij}}{\pl t_{\ell}}=\int_{\BR} z^{i+j+\ell}\,e^{\sum t_k z^k}\rho(dz)
.$$
Therefore the semi-infinite moment matrix
$m_{\iy}=(\mu_{ij})_{i,j\geq 0}$ satisfies
\be
\frac{\pl m_{\iy}}{\pl t_i}=\Lb^i m_{\iy}=m_{\iy}
 \Lb^{\top ^i}.
\ee
The point now is that the following integral can be expressed as a
determinant of moments, namely
\bean
n! \tau_n(t)=\int_{\HR_n}e^{-Tr\,V_t(X)}dX
 &=&\int_{\BR^n}\Delta^2(z)\prod^n_{k=1}\rho_t(dz_k)\\
 &=&\int_{\BR^n}\sum_{\sg\in S_n}\det
 \left(z^{\ell-1}_{\sg(k)}z^{k-1}_{\sg(k)}\right)_{1\leq\ell,k\leq
n}\prod^n_{k=1}\rho_t(dz_k)\\
 &=&\int_{\BR^n}\sum_{\sg\in
S_n}\det\left(z_{\sg(k)}^{\ell+k-2}\right)_{1\leq\ell,k\leq
n}\prod^n_{k=1}
\rho_t(dz_{\sg(k)})\\
&=&\sum_{\sg\in S_n}\det\left(\int_{\BR}
z_{\sg(k)}^{\ell+k-2}\rho_t(dz_{\sg(k)})
\right)_{1\leq\ell,k\leq n}\\
&=&n!\det\left(\int_{\BR}
z^{\ell+k-2}\rho_t(dz)\right)_{1\leq\ell,k\leq n}\\
&=&n!\det(\mu_{ij})_{0\leq i,j\leq n-1}
\eean
is a $\tau$-function for the KP-equation;
 also in view of (4.4) and the upper-lower Borel decomposition
 (0.3) of $m_{\iy}$,
 the integrals form a vector of $\tau$-functions for the Toda
 lattice.

\subsection{$\pl m / \pl t_k=\Lb^k m +m
\Lb^{\top k} $, skew-orthogonal polynomials and
symmetric and symplectic matrix integrals. }

\setcounter{equation}{4}

Consider a skew-symmetric semi-infinite matrix $$ m_{\iy}(t)
=(\mu_{ij}(t))_{i,j\geq 0},~~\mbox{with}~~m_n(t)=
 (\mu_{ij}(t))_{0\leq i,j\leq n-1},$$
satisfying \be
\pl m_{\iy} / \pl t_k=\Lb^k m_{\iy} +m_{\iy}
\Lb^{\top n} .\ee
Then we have shown in previous sections that, upon skew-Borel
decomposing $m_{\iy}$, these equations ultimately imply the
existence of functions $\tau(t)$ satisfying the bilinear equations
(3.4). Remember also
$$   h(t)=\mbox{diag}(\ldots,h_{-2},h_{-2}, h_0, h_0, h_2,
h_2,\ldots)\in
\DR_0^-,~~\mbox{with}~~ h_{2n}(t)=
\frac{ \tau_{2n+2}(t)}{ \tau_{2n}(t)}.$$
  Here, we need the Pfaffian $ pf (A)$ of a
  skew-symmetric matrix
$
A=(a_{ij})_{0\leq i,j\leq n-1}$ for\footnote{In the formula below
$
 (i_0,i_1,\ldots
,i_{n-2},i_{n-1})=\sigma (0,1,\ldots,n-1)$, where $\sigma$ is a
permutation and $\vr(\sigma) $ its parity.} even $n$:
\begin{eqnarray}
\lefteqn{pf(A)dx_0\wedge\cdots\wedge dx_{n-1}}\nonumber\\
&=&\frac{1}{n!}\left(
\sum_{0\leq i<j\leq n-1}
a_{ij}dx_i\wedge dx_j\right)^n\nonumber\\
&=&\frac{1}{2^{n/2}(n/2)!}\left(\sum_{\sg}\vr(\sg)a_{i_0,i_1}a_{i_2,i_3}
\cdots a_{i_{n-2},i_{n-1}}
\right)dx_0\wedge \cdots\wedge dx_{n-1},\nonumber\\
\end{eqnarray}
so that $pf(A)^2=\det A$. We now state the following theorem due to
Adler-Horozov-van Moerbeke\cite{AHV}.

\begin{theorem}
Consider a semi-infinite skew-symmetric matrix $m_{\iy}$, evolving
according to (4.5); setting
\be
\tau_{2n}(t)=pf (m_{2n}(t)),~\mbox{and}~~ h_{2n}=
\frac{pf (m_{2n+2}(t))}{pf (m_{2n}(t))},
\ee
 then the wave vector $\Psi_1$, defined by (3.3) is a sequence of polynomials, except for the
exponential,
\be
\Psi_{1,k}(t,z)=e^{\sum t_i z^i} q_k(t,z),
\ee
where the $ q_k$'s are skew-orthonormal polynomials of the form
(0.17) satisfying
\be
(\la  q_i,  q_j\ra^{sk})_{0\leq i,j<\iy}
 =J, ~~\mbox{with}~~\la  y^i,  z^j\ra^{sk}:=\mu_{ij}.
\ee
The matrix $Q$ defined by $q(z)=Q\chi(z)$ is the unique solution
(modulo signs)
 to the skew-Borel decomposition of $m_{\iy}$:
\be
m_{\iy}(t)=Q^{-1}JQ^{\top -1}, ~~\mbox{with}~~Q\in \Bk. \ee The
matrix $L=Q\Lb Q^{-1}$, also defined by $$z q(t,z)=L q(t,z),$$
 and the diagonal matrix $ h$ satisfy the equations
\be
\frac{\pl L}{\pl t_i}=\left[-\pi_{\Bk}L^i, L \right].
~~\mbox{and}~~
 h^{-1}\frac{\pl  h}{\pl t_i}
=2\pi_{\Bk} (L^i)_0  .
\ee
\end{theorem}

\underline{\sl Sketch of proof}: at first note that looking
 for skew-orthogonal polynomials is tantamount to the
 skew-Borel decomposition of $m_{\iy}$, so that
 (4.9) and (4.10) are equivalent. The skew-orthogonality of
 the polynomials (0.17) follows from expanding the determinants
 explicitly in terms of $z$-columns, upon using the expression
 for the pfaffian in terms of a column
 $$
\sum_{0\leq k\leq\ell-1}(-1)^k a_{ ki}
pf(0,\dots,\hat k,\dots,\ell-1)=pf(0,\dots,\ell-1,i).
$$
For details, see \cite{AHV}. On the other hand, Theorem 3.2 gives
$\Psi(t,z)$ and hence $Q$ in terms of $\tau_n(t)$ of
 (4.7). By the uniqueness of the decomposition (4.10),
 the two ways of arriving at $Q$, (0.16) and (3,3) must coincide.\qed

\vspace{1.2cm}

\noindent \underline{\sl Important remark}: The polynomials
 (0.16) provide an explicit algorithm to perform the skew-Borel
 decomposition of the skew-symmetric matrix $m_{\iy}$. Namely,
 the coefficients of the polynomials $q_i$ provide the entries
 of the matrix $Q$. This fact will be used later in the examples.

\vspace{2cm}

\noindent{\bf Symmetric matrix integrals}
Here we shall focus on integrals of the type
 \be
 \int_{{\cal S}_{2n}}
 e^{Tr~(- V(X)+\sum_1^{\iy} t_i X^i )} dX,
 \ee
where $dX$ denotes Haar measure
\be
dX:=\displaystyle{\prod_{1\leq i\leq j\leq n}}d\Re X_{ij}=|\Delta
(z)|dz_1\cdots dz_n~dU,
\ee
over the space $S_{2n}$ of symmetric matrices. As will appear
below, the integral (4.12) leads to
a skew-inner-product with weight
$\rho_t(z)dz:=e^{-V_t(z)}dz:=e^{-V(z)+\sum t_i z^i}dz=e^{\sum t_i
z^i}\rho(z)dz$ on an interval $\subseteq\BR$, as in (0.1),
\be
\la f(x),g(y)\ra:= \int\!\int_{\BR^2}f(x)g(y) \vr(x-y) \rho_t(dx)\rho_t(dy)
\ee
and therefore skew-symmetric moments\footnote{$\vr(x)=1$, for
$x\geq 0$ and $=-1$, for $x<0$.}
\bea
\mu_{ij}(t)&=&\int\!\int_{\BR^2}x^iy^j\vr(x-y)
\rho_t(x)\rho_t(y)dxdy
\nonumber\\
&=&\int\!\int_{x\geq y}(x^iy^j-x^jy^i)\rho_t(x)\rho_t(y)dxdy
\nonumber\\ &=&\int_{\BR}\left(F_j(x)G_i(x)-F_i(x)G_j(x)\right)dx.
\eea
where
$$ F_i(x):=\int^x_{-\iy}y^i\rho_t(y)dy
\quad\mbox{and}
\quad G_i(x):=F_i'(x)=x^i\rho_t(x).
$$
By simple inspection, the moments
 $\mu_{k\ell}(t)$ satisfy
\bean
\frac{\pl \mu_{k\ell}}{\pl
t_i}&=&\int\!\int_{\BR^2}(x^{k+i}y^{\ell}+x^ky^{\ell
+i})\vr(x-y)\rho_t(x)\rho_t(y)dx dy\\
&=&\mu_{k+i,\ell}+\mu_{k,\ell+i},
\eean
and so $m_{\iy}$ satisfies (4.5).

According to Mehta \cite{M}, the symmetric matrix integral can now
be expressed in terms of the pfaffian, as follows, taking into
account a constant $c_{2n}$, coming from integrating the orthogonal
group:

\bean
\lefteqn{\frac{1}{(2n)!}\int_{{\cal S}_{2n}(E)}
e^{Tr~( -V(X)+\sum t_i X^i) } dX}
\\
&=&\frac{1}{(2n)!}\int_{\BR^{2n}}|\Delta_{2n}(z)|
\prod_{i=1}^{2n}
\rho_t(z_i)dz_i\\
&=&\int_{-\iy<z_1<z_2<\cdots<z_{2n}<\iy}\det\left(z_{j+1}^i
\rho_t(z_{j+1})\right)_{0\leq i,j\leq 2n-1}\prod_{i=1}^{2n}
dz_i,\\ &=&\int_{-\iy<z_2<z_4<\cdots<z_{2n}<\iy}
\prod^n_{k=1}\rho_t(z_{2k})dz_{2k} \\ & &
\det\left(\int_{-\iy}^{z_2}z_1^i\rho_t(z_1)dz_1~,~z_2^i~,\dots,~
\int_{z_{2n-2}}^{z_{2n}}z_{2n-1}^i\rho_t(z_{2n-1})dz_{2n-1}~,
~z_{2n}^i
\right)_{0\leq i\leq 2n-1}\\
&=&\int_{-\iy<z_2<z_4<\cdots<z_{2n}<\iy}
\prod^n_{k=1}\rho_t(z_{2k})dz_{2k}\\ & &
\det\left(F_i(z_2)~,~z_2^i~,~F_i(z_4)-F_i(z_2)~,~
z_4^i~,~\dots~,~F_i(z_{2n})-F_i(z_{2n-2})~,~z^i_{2n}\right)_{0\leq
i\leq 2n-1}\\
 &=&\int_{-\iy<z_2<z_4<\cdots<z_{2n}<\iy}\prod^n_1dz_i~~
\det\Bigl(F_i(z_2)~,~G_i(z_2)~,~\dots,~F_i(z_{2n})~,~
G_i(z_{2n})\Bigr)_{0\leq i\leq 2n-1},\\
 &=&\frac{1}{n!}\int_{\BR^n}\prod^n_1dy_i~~
\det\Bigl(F_i(y_1)~,~G_i(y_1)~,~\dots,~F_i(y_n)~,~
G_i(y_n)\Bigr)_{0\leq i\leq 2n-1},\\
&=&{\det}^{1/2}\left(\int_{\BR}
 (G_i(y)F_j(y)-F_i(y) G_j(y))dy\right)_{0\leq
i,j\leq 2n-1}\\
 & &\mbox{\hspace{5cm} using de
Bruijn's Lemma \cite{M},p.446},\\ &=& pf
 \left(\mu_{ij}\right)_{0\leq
i,j\leq 2n-1} \\ &=& \tau_{2n}(t),
\eean
which is a Pfaffian $\tau$-function.

\noindent{\bf Symplectic matrix integrals}: Here we shall concentrate on integrals of the type
 \be
 \int_{{\cal T}_{2n}}
 e^{2Tr~(- V(X)+\sum_1^{\iy} t_i X^i )} dX,
 \ee
where $dX$ denotes Haar measure\footnote{$\bar X$ means the usual complex
conjugate. The condition on the $2\times 2 $ matrices $X_{k \ell}$
implies that $X_{kk}=X_k I$, with $X_k\in
\BR$ and $I$ the identity.}
$$
dX=\prod^N_1 dX_k \prod_{k \leq \ell}
 dX_{k \ell}^{(0)}\bar {dX_{k \ell}^{(0)}}
   dX_{k \ell}^{(1)} \bar {dX_{k \ell}^{(1)}},
 $$
 on the space $
\TR_{2N}$ of self-dual $N\times N$ Hermitean matrices, with quaternionic
entries; the latter can be realized as the space of $2N
\times 2N$ matrices with entries $X^{(i)}_{
\ell k}\in \BC$
$$
\TR_{2N}=\left\{ X=(X_{k \ell})_{1\leq k ,\ell \leq N} ,
X_{k \ell}=\pmatrix{X^{(0)}_{k \ell}&X^{(1)}_{k \ell}\\ \cr
        -\bar X^{(1)}_{k \ell}&\bar X^{(0)}_{k \ell}  }
        \mbox{ with }~X_{ \ell k}=\bar X_{k \ell}^{\top}
\right\},
$$

A more exotic
  skew-symmetric matrix $m_{\iy}$ satisfying (4.5) is given by the
  moments, with $V(y,t)=e^{2 (-V(y)+\sum t_{\al}y^{\al})}$,
\begin{eqnarray}
  \mu_{ij}(t)&=&\int_{\BR}
 \{y^i,y^j\}e^{2 (-V(y)+\sum t_{\al}y^{\al})}I_E(y)dy\nonumber\\
 &=& \int_{\BR}
 \{y^i e^{ V(y,t)},y^j e^{ V(y,t)}\} I_E(y)dy\nonumber\\
 &=& \int_{\BR}
 (G_i(y)F_j(y)-F_i(y) G_j(y))dy,
 \end{eqnarray}
upon setting
$$
 F_j(x)=x^j e^{V(x,t)} ~~\mbox{and}~~
 G_j(x):=F'_j(x)=\left(x^j e^{V(x,t)}\right)' .
$$
That $m_{\iy}$ satisfies (4.5) follows at once from the
 first expression (4.17) above.

\bean
\mu_{k\ell}(t)&=&\int\{y^k,y^{\ell}\}\rho_t(y)^2 dy\\
&=&\int(k-\ell)y^{k+\ell-1}\rho_t(y)^2  dy\\
 \frac{\pl\mu_{k\ell}}{\pl t_i}&=&2\int\{y^k,y^{\ell}\}
 y^ie^{2(-V(y)+\sum t_iy^i)}dy\\
 &=&\int\left((k+i-\ell)y^{k+i+\ell
-1}+(k-\ell-i)y^{k+i+\ell -1}\right) \rho_t(y)^2  dy\\
&=&\mu_{k+i,\ell}+\mu_{k,\ell +i},
\eean
thus leading to (4.5). Using the relation
$$
\prod_{1\leq i,j\leq n}(x_i-x_j)^4=\det\left(x_1^i~~~(x_1^i)'~~~x_2^i~~~(x_2^i)'~\dots~
~x_n^i~~~(x_n^i)'\right)_{0\leq i\leq 2n-1},
$$
one computes, using again de Bruijn's Lemma,

\noindent $\displaystyle{\frac{1}{(n)!}
\int_{{\cal T}_{2n}}e^{2~Tr
 (-V(X)+\sum t_i X^i )} dX}$
 \bean
&=&\frac{1}{n!}\int_{\BR^n}\prod_{1\leq i,j\leq n}(x_i-x_j)^4
  \prod_{i=1}^{n}\left(e^{-2V(x,t)} dx_i   \right)\\
 &=&\frac{1}{n!}\int_{\BR^n}
\prod^n_{k=1}\left(dx_{k} e^{-2V(x_{k},t)}dx_k
 \right)\\ & &
\hspace{3cm}\det\left(x_1^i~~~(x_1^i)'~~~x_2^i~~~(x_2^i)'~\dots~
~x_n^i~~~(x_n^i)'\right)_{0\leq i\leq 2n-1}\\
&=&\frac{1}{n!}\int_{\BR^n}\prod^n_1dy_i~~
\det\Bigl(F_i(y_1)~~G_i(y_1)~~\dots~~F_i(y_n)~~
G_i(y_n)\Bigr)_{0\leq i\leq 2n-1},\\
&=& {\det}^{1/2} \left(\int_{\BR}
 (G_i(y)F_j(y)-F_i(y) G_j(y))dy\right)_{0\leq
i,j\leq 2n-1}\\ &=&pf \left(\mu_{ij}\right)_{0\leq i,j\leq 2n-1}\\
\\ &=& \tau_{2n(t)},\eean
which is a Pfaffian $\tau$-function as well.

\section{A map from the Toda to the Pfaff lattice}

Remember from (0.1), the notations $\rho_t(z)=\rho (z)
 e^{\sum t_k z^k}$,
and $\rho'/\rho =-g/f$.
  Assuming, in addition, $f(z)\rho(z)$ vanishes at the endpoints
  of the interval under consideration (which could be finite, infinite
  or semi-infinite), one checks the the $t$-dependent
  operator in $z$,
\bea
\Bn_t:&=&\sqrt{\frac{f}{\rho_t}}\frac{d}{dz}
\sqrt{f\rho_t}\nonumber\\
&=&e^{-\frac{1}{2}\sum t_k
z^k}\left(\frac{d}{dz}f(z)-\frac{f'+g}{2}(z)\right)
e^{\frac{1}{2}\sum t_k z^k}\nonumber\\ &=&
 \frac{d}{dz}f(z)-\frac{f'+g_t}{2}(z),~~\mbox{with}~
 g_t(z)=g(z)-f(z)\sum_1^{\iy}kt_k z^{k-1},\nonumber\\
\eea
maintains $\HR_+=\{1,z,z^2,...\} $ and is skew-symmetric with
respect to the
 $t$-dependent inner-product $\la ~,\ra_t^{sy}$, defined by
 the weight $\rho_t(z)dz$,
 $$
 \la \Bn_t\varphi,\psi\ra_t^{sy}
 =\int_E (\Bn_t\varphi )(z)\psi(z)\rho_t(z) dz=
 -\int_E \varphi (\Bn_t\psi)\rho_t dz
         =-\la \varphi, \Bn_t\psi\ra_t^{sy}.
 $$
 The orthonormality of the $t$-dependent
  polynomials $p_n(t,z)$ in $z$ imply
$$\la p_n(t,z), p_m(t,z)\ra_t^{sy}=\dt_{mn}.$$
 The matrices $L$ and $M$ are defined by
 $$
 zp=Lp~~~\mbox{and}~~~
 e^{-\frac{1}{2}\sum t_k z^k}\frac{d}{dz}
 e^{\frac{1}{2}\sum t_k z^k}p=Mp.
 $$
 The skewness of $\Bn_t$ implies the skew-symmetry
 of the matrix
 \be
\NR(t)=f(L)M-\frac{f'+g}{2}(L),~~\mbox{such that} ~~\Bn_t p(t,z)
=\NR p(t,z);
\ee
so $\NR(t)$ can be viewed as the operator $\Bn_t$, expressed in the
polynomial basis $(p_0(t,z),p_1(t,z),...)$.

In the next theorem, we shall
 consider functions $F$ of two (non-commutative)
 variables $z$ and $\Bn_t$ so that the (pseudo)-differential
 operator in $z$ and the matrix
$$\Bu_t:=F(z,\Bn_t)~~\mbox{and}~~\UR:=F(L,\NR), $$
related by\footnote{with the understanding that $F(L,\NR)$
 reverses the order of $z, \Bu$ in $F(z,\Bu)$.}
$$F(z,\Bn_t)p(t,z)= F(L,\NR)p(t,z),$$ are skew-symmetric as well. Examples of
$F$'s are\footnote{$\{A,B\}^{\dag}=AB+BA.$}
$$F(z,\Bn_t):=\Bn_t,~\Bn_t^{-1} ~\mbox{or}~
\{z^{\ell},\Bn_t^{2k+1}\}^{\dag},
$$ corresponding to $$F(L,\NR)=
\NR,~~\NR^{-1}~\mbox{or}~~
\{\NR^{2k+1},L^{\ell}\}^{\dag}.$$

\begin{theorem}
Any H\"ankel matrix $m_{\iy}$ evolving according to the vector
fields
$$
\frac{\pl m_{\iy}(t)}{\pl t_k}=\Lb^k m_{\iy}
$$
leads to matrices $L$ and $M$, evolving according to the Toda
lattice (1.9). Consider a function $F$ of two variables,
 such that the operator $\Bu_t:=F(z,\Bn_t)$ is skew-symmetric
 with respect to $\la ~,\ra_t^{sy}$ and so the matrix
 $$\UR(t)=F(L(t),\NR(t)),~~
 \mbox{defined by}~~\Bu_t p(t,z)=\UR p(t,z)
 $$ is skew-symmetric. This induces a natural
 lower-triangular matrix $O(t)$, mapping the Toda
 lattice into the Pfaff lattice:

\vspace{.5cm}
$$
\mbox{Toda lattice}\left\{\begin{tabular}{l}
$p_n(t,z)=\left(S(t)\chi(z)\right)_n ~~
 \mbox{orthonormal with respect to} $\\
 $\hspace{4cm}~m_{\iy}(t)=\left( \la z^i,z^j\ra^{sy}_t
 \right)_{0\leq i,j \leq \iy}=S^{-1}S^{\top -1}$
 \\
 \\
 $L(t)=S\Lb S^{-1} ~~\mbox{satisfies}~\quad\displaystyle{\frac{\pl L}{\pl t_j}
 =\left[-\frac{1}{2}\pi_{bo} L^j,
L\right],~j=1,2,...}$\\
\end{tabular}
\right.
$$
$$\quad\left\downarrow
\vbox to1.1in{\vss}\right.
\mbox{ map $O(2t)$ such that }
\left\{
\begin{array}{l}
  -\UR(2t)= O^{-1}(2t)JO^{\top -1}(2t)\\[6pt]
  O(2t) \mbox{ is lower-triangular } \\[6pt]
   O(2t)S(2t) \in \GR_{\Bk}
\end{array}
\right.
.
$$
$$
\mbox{Pfaff lattice}\left\{\begin{tabular}{l}
 $q_n(t,z)=
 \left(O(2t)p(2t,z)\right)_n,
 \mbox{skew-orthonormal with regard to }$\\
\\
 \hspace{1.4cm}$
 \tilde m_{\iy}(t):=-S^{-1}(2t)\UR(2t) S^{\top
-1}(2t)=Q^{-1}(t)JQ^{\top -1}(t)$\\  \\
$  \hspace{2.8cm}=\left( \la z^i, z^j\ra^{sk}_t \right)
_{0\leq i,j \leq \iy} $\\ \\
 \hspace{2.7cm} $=\left( \la z^i, \Bu_{2t}z^j\ra^{sy}_{2t} \right)
_{0\leq i,j \leq \iy}
$\\
 \\
 \\
 $\tilde L(t):=O(2t)L(2t)O(2t)^{-1}~~\mbox{satisfies} ~
\displaystyle{\frac{\pl \tilde L}{\pl t_j}}=[-\pi_{\Bk}\tilde
L^j,\tilde L]~,j=1,...$\\
\end{tabular}
\right.
$$
\end{theorem}

\proof Since $\UR(t)$ is skew-symmetric, it admits a
 skew-Borel decomposition
\be
-\UR(t)=O^{-1}(t)JO^{\top -1}(t),~~\mbox{with lower-triangular }O(t)
.\ee
But the new matrix, defined by
\be
\tilde m_{\iy}(t):=-S^{-1}(2t)\UR(2t)S^{\top -1}(2t),
\ee
is skew-symmetric and thus admits a unique skew-Borel decomposition
\be
\tilde m_{\iy}(t)=\tilde Q^{-1}(t)J\tilde
Q(t)^{\top -1}~~\mbox{with}~~\tilde Q(t)\in \GR_{\Bk}.
\ee
Comparing (5.3), (5.4) and (5.5) leads to a unique choice of matrix
$O(t)$, skew-Borel decomposing $-\UR(2t)$, as in (5.3), such that
\be
O(2t)S(2t)=\tilde Q(t) \in \GR_{\Bk}.
\ee
Using
$$
\frac{\pl \UR}{\pl t_k}(2t)=[\pi_{sy}L^k(2t),\UR(2t)]
$$
and
$$
\frac{\pl S}{\pl t_k}(2t)=-(\pi_{bo} L^k(2t))S(2t),
$$
we compute

\medbreak

\noindent $\displaystyle{\frac{\pl\tilde m_{\iy}}{\pl t_k}(t)}$
\begin{eqnarray*}
&=&S^{-1}\frac{\pl S}{\pl t_k}(2t)S^{-1}\UR(2t)S^{\top
-1}(2t)-S^{-1}(2t)\left(\frac{\pl}{\pl t_k}\UR(2t)\right)S^{\top
-1}(2t)\\ & &\hspace{2cm}+S^{-1}(2t)\UR(2t)S^{\top -1}\frac{\pl
S^{\top}}{\pl t_k} (2t)S^{\top -1}\\ &=&-S^{-1}(\pi_{bo} L^k(2t))
\UR S^{\top
-1}-S^{-1}[\pi_{sy}L^k,\UR]S^{\top -1}-
S^{-1}\UR(\pi_{bo} L^k)^{\top}S^{\top -1}\\
 &=&-S^{
-1}(\pi_{bo}L^k+\pi_{sy} L^k)\UR S^{\top
-1}-S^{-1}\UR ((\pi_{bo}L^k)^{\top}-\pi_{sy}L^k)S^{\top -1}\\
&=&-S^{ -1}L^k \UR S^{\top -1}-S^{ -1}\UR L^{\top k}S^{\top
-1},\,\,\mbox{using (5.6) below}\\ &=&-\Lb^kS^{ -1}\UR S^{\top -1}-S^{
-1}\UR S^{\top -1}\Lb^{\top k}S^{\top}S^{\top
-1},\,\,\mbox{using\,\,}L^k=S\Lb^kS^{ -1},\\
&=&\Lb^k\tilde m_{\iy}(t)+\tilde m_{\iy}(t)\Lb^{\top k}.
\end{eqnarray*}

For an arbitrary matrix $A$, we have
\be A=A^{\top}\Longleftrightarrow
A=(A_{bo})^{\top}-A_{sy}.
\ee
Indeed, remembering that\footnote{$A_{\pm}$ means the usual
strictly upper(lower)-triangular part and $A_0$ the diagonal part
in the common sense.}$A_{bo}=2A_-+A_0$ and $A_{sy}=A_+-A_-$, one
checks
$$
(A_{bo})^{\top}-A_{sy}-A=
2(A_-)^{\top}+A_0-(A_+-A_-)-A_--A_+-A_0=-2(A_+-(A_-)^{\top}).
$$
so that the left hand side vanishes, if the right hand side does;
the latter means A is symmetric.

We now define $\tilde L(t)$ by conjugation of $L(2t)$ by $O(2t)$:
$$
\tilde L(t):=O(2t)L(2t)O(2t)^{-1}=O(2t)S(2t)\Lb
S^{-1}(2t)O(2t)^{-1}=\tilde Q(t)\Lb\tilde Q^{-1}(t).
$$
Therefore the sequence of polynomials
$$
q(t,z):=O(2t)p(2t,z)=O(2t)S(2t)\chi(z)=\tilde Q(t)\chi(z)
$$
is skew-orthonormal
$$
\la q_i(t,z),q_j(t,z)\ra^{sk}=J_{ij}
$$
with regard to the skew inner-product specified by the matrix $\tilde m_{\iy}$:
$$
\la z^i,z^j\ra_t^{sk}=\tilde\mu_{ij}(t).
$$
In the last step, we show that
 $\la\vp,\psi\ra^{sk}=\la\vp,\Bu\psi\ra^{sy}$. Since
\begin{eqnarray}
\UR(2t)&=&-O^{-1}(2t)JO^{\top
-1}(2t)  ,
\end{eqnarray}
we compute
\begin{eqnarray}
 \la q_i(t,z),(\Bu_{2t} q)_j(t,z)\ra_{2t}^{sy}&=&\left\la (Op)_i(2t),
 (\Bu Op)_j(2t)\right\ra_{2t}^{sy}\nonumber\\
 &=&\la (Op)_i(2t),(O\Bu p)_j(2t)\ra_{2t}^{sy}\nonumber\\
 &=&\la (Op)_i(2t),(O \UR p)_j(2t)\ra_{2t}^{sy}\nonumber
 \\ &=&(O(2t)\left\la p_k(2t),p_{\ell}(2t)\right\ra^{sy}_{k,\ell\geq
 0}(O \UR )^{\top}(2t))_{ij}\nonumber
 \\ &=&(O(2t)I(O \UR )^{\top}(2t))_{ij}\nonumber\\
 &=&(O(2t)\UR^{\top}(2t)O^{\top}(2t))_{ij}\nonumber\\
 &=&-(O(2t)\UR (2t)O^{\top}(2t))_{ij}\nonumber\\ &=&J_{ij}
 ,~\mbox{using} ~(5.8).
\end{eqnarray}
Therefore, defining a new skew inner-product $\la \, ,\,\ra^{
 sk^{\prime}}$
$$
\la\vp,\psi\ra^{sk^{\prime}}:=\la\vp,\Bu \psi\ra^{sy},
$$
we have shown
$$
\la q_i,q_j\ra_t^{sk^{\prime}}=\la q_i,q_j\ra_t^{sk}=J_{ij},
$$
and so by completeness of the basis $q_i$, we have
$$
\la ~,~\ra_t^{sk^{\prime}}=\la ~,~\ra_t^{sk},
$$
thus ending the proof of Theorem 4.1.
\qed

\medbreak

\section{Example 1: From Hermitean to symmetric matrix integrals}

Striking examples are given by using the map $O(t)$ obtained from
skew-borel decomposing $\NR^{-1}(t)$ and $\NR(t)$; see (5.2). This
section deals with $\NR^{-1}(t)$, whereas the next will deal with
$\NR(t)$.

\begin{proposition}
  The special transformation
$$\UR(t)=\NR^{-1}(t)=\left(f(L)M-\frac{f'+g}{2}(L)\right)^{-1}(2t)$$
 maps the Toda lattice $\tau$-functions with initial weight\footnote{Remember
 $\rho'/\rho=-V'=-g/f$.}
$\rho=e^{-V},~V'=-g/f$ (Hermitean matrix integral) to the Pfaff
lattice $\tau$-functions (symmetric matrix integral), with initial
weight
 $$\tilde \rho_t(z):=\left( \frac{\rho_{2t}(z)}{f(z)} \right)^{\frac{1}{2}}
 =e^{-\frac{1}{2}(V(z)+\log
f(z)-2\sum_1^{\iy} t_i z^i)}=:e^{-\tilde V(z)+\sum_1^{\iy} t_i
z^i}.$$ To be precise:

$$
\mbox{Toda lattice}\left\{\begin{tabular}{l}
$p_n(t,z) ~~\mbox{orthonormal polynomials in $z$
 for the inner-product
}$ \\\hspace{4cm} $\la \varphi,\psi\ra_t^{sy}=
=\displaystyle{\int \varphi(z)\psi(z)
 \rho_t(z)dz} $,\\
$\mu_{ij}(t)=\la z^i , z^j \ra^{sy}_t $ and
 $ m_n=(\mu_{ij})_{0\leq i,j\leq n-1},$\\ $\tau_n(t)=\det
m_n=\displaystyle{\frac{1}{n!}\int_{\HR_n}
e^{Tr(-V(X)+\sum_1^{\iy}t_iX^i)}dX}$
\\

\end{tabular}
\right.
$$
$$\quad\left\downarrow\vbox to1.1in{\vss}\right.
\mbox{map $O(2t)$ such that}\
\left\{
\begin{array}{l}
  -\NR^{-1}(2t)= O^{-1}(2t)JO^{\top -1}(2t)\\[6pt]
  O(2t) \mbox{ is lower-triangular } \\[6pt]
   O(2t)S(2t) \in \GR_{\Bk}
\end{array}
\right.
$$
$$
\mbox{Pfaff lattice}\left\{\begin{tabular}{l}
 $q_n(t,z)=O (2t) p_n(2t,z) ~~\mbox{skew-orthonormal
  polynomials}$\\
\hspace{3.0cm}$\mbox{ in
$z$ for the skew-inner-product (weight $\tilde \rho$}$),
 \\

\\
 \hspace{1cm}$
 \la \varphi,\psi \ra_t^{sk}
 :=\la \varphi,\Bn_{2t}^{-1}\psi \ra_{2t}^{sy}$\\  \\
$  \hspace{2.6cm}=\displaystyle{\frac{1}{2}
 \int\!\int_{\BR^2}\varphi(x) \psi(y)
 \vr(x-y) \tilde\rho_t(x) \tilde\rho_t(y)dx\,dy}
 $\\ \\

  $\tilde
\mu_{ij}(t)=\la x^i, y^j\ra ^{sk}_t$ and
 $\tilde m_n=(\tilde\mu_{ij})_{0\leq i,j\leq n-1}$
 \\
 \\
$\tilde\tau_{2n}(t)=pf(\tilde
m_{2n})=\displaystyle{\frac{1}{(2n)!}\int_{{\cal
S}_{2n}}e^{Tr(-\tilde V(X)+\sum_1^{\iy}t_iX^i)}dX}.$

\end{tabular}
\right.
$$
\end{proposition}

\bigbreak

In the first integral defining $\tau_n(t)$, $dX$ denotes Haar
measure on Hermitean matrices (see section 4.1), whereas the second
integral $\tilde\tau_{2n}(t)$ involves Haar measure on symmetric
matrices (see section 4.2)

\proof At first, check that
\be
\left(\frac{d}{dx}\right)^{-1}\vp(x)=\frac{1}{2}\int\vr(x-y)\vp(y)dy.
\ee
Indeed,
\begin{eqnarray*}
\frac{d}{dx}\left(\frac{d}{dx}\right)^{-1}\vp(x)&=&\int\frac{1}{2}\frac{\pl}{\pl x}
\vr(x-y)\vp(y)dy\\
&=&\int\delta(x-y)\vp(y)dy~~\mbox{using}~~\frac{\pl}{\pl x}\vr(x)=2
\dt(x)\\ &=&\vp(x).
\end{eqnarray*}
Consider now the operator
$$
\Bu_t=\Bn_t^{-1}=\left(\sqrt{\frac{f}{\rho_t}}\frac{d}{dz}
 \sqrt{ f \rho_t}\right)^{-1},~\mbox{so that }~
  \Bu_tp=\Bn_t^{-1}p=\NR^{-1}p ,
$$
according to (5.2). Let it act on a function $\vp(x)$:
\begin{eqnarray*}
\Bn_t^{-1}\vp (x)
&=&\left(\frac{1}{\sqrt{f(x)\rho_t(x)}}\left(\frac{d}{dx}\right)^{-1}
\sqrt{\frac{\rho_t(x)}{f(x)}}\right)\vp(x)\\
&=&\int_{\BR}\frac{1}{\sqrt{f(x)\rho_t(x)}}\frac{\vr(x-y)}{2}
\sqrt{\frac{\rho_t(y)}{f(y)}}\vp(y)dy,
\,\,\mbox{using (6.1)}.
\end{eqnarray*}
One computes
\begin{eqnarray*}
\la\vp,\psi\ra_t^{sk}
 &=&\la\vp,\Bu_{2t}\psi\ra_{2t}^{sy}\\
 &=&\la\vp,\Bn^{-1}_{2t}\psi\ra_{2t}^{sy}\\
&=&\frac{1}{2}\int\!\int_{\BR^2}\sqrt{\frac{\rho_{2t}(x)}{f(x)}}
 \vr(x-y) \sqrt{\frac{\rho_{2t}(y)}{f(y)}}\vp(x)\psi(y)dx\,dy\\
&=&\frac{1}{2}\int\!\int_{\BR^2}\tilde\rho(x) \tilde\rho(y)
 e^{\sum_1^{\iy} t_k(x^k+y^k)}\vr(x-y)\vp(x)\psi(y)dx\,dy.
\end{eqnarray*}
So, finally setting $\tilde V(x)=\frac{1}{2}(V(x)+\log f(x))$
yields
$$
\tilde\tau_{2n}(t)=pf(\tilde m_{2n})=\frac{1}{(2n)!}\int_{{\cal S}_{2n}}e^{Tr(-\tilde
V(X)+\sum_1^{\iy}t_iX^i)}dX.
$$

\bigbreak

\noindent{\bf The map $O$ for the
classical orthogonal polynomials at $t=0$}: {\em Then, the matrix
$O$, mapping orthonormal $p_k$ into skew-orthonormal polynomials
$q_k$, is given by a lower-triangular three-step relation:
\bea
 q_{2n}(0,z)&=&\sqrt{\frac{c_{2n}}{a_{2n}}}p_{2n}(0,z)\nonumber\\
 q_{2n+1}(0,z)&=&\sqrt{\frac{a_{2n}}{c_{2n}}}\nonumber\\
&&
\left(-c_{2n-1}p_{2n-1}(0,z)+\frac{c_{2n}}{a_{2n}}(\sum_0^{2n}b_i)
 p_{2n}(0,z)+c_{2n}p_{2n+1}(0,z)\right)\nonumber\\
\eea
where the $a_i$ and $b_i$ are the entries in the tridiagonal matrix
defining the orthonormal polynomials, and the $c_i$'s are the
entries of the skew-symmetric matrix $\NR$.}

In \cite{AvM2}, we showed that then $\NR$ is tridiagonal, at the
same time as $L$, (see Appendix 2)
\be
L=\left[\begin{array}{ccccc}
b_0&a_0& & & \\
a_0&b_1&a_1& & \\
 &a_1&b_2&\ddots&\\
 & &\ddots&\ddots&
\end{array}
\right],\quad -\NR=\left[\begin{array}{ccccc}
0&c_0& & & \\
-c_0&0&c_1& & \\
 &-c_1&0&\ddots&\\
 & &\ddots& &
\end{array}\right]\,,
\ee
 with the following precise entries:

\noindent \underline{Hermite}: $\rho(z)=e^{-z^2}$,
$a_{n-1}=\sqrt{\frac{n}{2}}$, $b_n=0$, $c_n=a_n$

\noindent \underline{Laguerre}: $\rho(z)=e^{-z}z^{\al}I_{[0,\iy)}(z)$, $a_{n-1}=\sqrt{n(n+\al)}$,
$b_n=2n+\al+1$, $c_n=a_n /2$

\vspace{0.4cm}

\noindent \underline{Jacobi}: $\rho(z)=(1-z)^{\al}(1+z)^{\rho}I_{[-1,1]}(z)$

\bea
a_{n-1}&=&\left(\frac{4n(n+\al+\beta)(n+\al)(n+\beta)}{(2n+\al+\beta)^2(
2n+\al+\beta+1)(2n+\al+\beta-1)}\right)^{1/2}\nonumber\\
b_n&=&\frac{\al^2-\beta^2}{(2n+\al+\beta)(2n+\al+\beta-2)}\nonumber\\
 c_n&=&a_n\left( \frac{\alpha+\beta}{2}+n+1 \right)\nonumber
 \eea
 If the skew-symmetric matrix $\NR$ has the
tridiagonal form above, then one checks its inverse has the
following form:
\be
-\NR^{-1}=\left(\begin{array}{ccccccccc}
0&-\frac{1}{c_0}&0&\frac{-c_1}{c_0c_2}&0&
\frac{-c_1c_3}{c_0c_2c_4}&0&
\frac{-c_1c_3c_5}{c_0c_2c_4c_6}&\\
\frac{1}{c_0}&0&0&0&0&0&0&0&\\
0&0&0&-\frac{1}{c_2}&0&\frac{-c_3}{c_2c_4}&0&\frac{-c_3c_5}{c_2c_4c_6}\\
\frac{c_1}{c_0c_2}&0&\frac{1}{c_2}&0&0&0&0& & \\
0&0&0&0&0&-\frac{1}{c_4}&0&\frac{-c_5}{c_4c_6} &\\
\frac{c_1c_3}{c_0c_2c_4}&0&\frac{c_3}{c_2c_4}&0&\frac{1}{c_4}&0&0&0\\
0&0&0&0&0&0&0&-\frac{1}{c_6}& \\
\frac{c_1c_3c_5}{c_0c_2c_4c_6}&0&\frac{c_3c_5}{c_2c_4c_6}
&0&\frac{c_5}{c_4c_6}&0&\frac{1}{c_6}&0 &\\ &&&&&&&&\ddots
\end{array}
\right).
\ee
In order to find the matrix $O$, we must perform the skew-Borel
decomposition of the matrix $-\UR$
$$-\UR=-\NR^{-1}=O^{-1}JO^{\top -1}.
$$
The recipe for doing so was given in theorem 4.1 (see also the
important remark, following that theorem). It suffices to form the
pfaffians (0.17), by appropriately bordering the matrix
$-\NR^{-1}$, as in (0.17), with rows and columns of powers of $z$,
yielding monic skew-orthogonal polynomials; we choose to call them
$r$'s, instead of the $q$'s of theorem 4.1,
 with $O\chi(z)=r(z)$.
They turn out to be the following simple polynomials, with
$1/\tilde{\tilde\tau}_{2n}=c_0c_2c_4\cdots c_{2n-2}$
\bea
 r_{2n}(z)&=&\frac{1}{\sqrt{\tilde{\tilde\tau}_{2n}\tilde{\tilde\tau}_{2n+2}}}
 \frac{c_{2n}z^{2n}}{c_0c_2\cdots c_{2n}}=
 \frac{1}{\sqrt{c_{2n}}}c_{2n}z^{2n}\nonumber\\
 r_{2n+1}(z)&=&
\frac{1}{\sqrt{\tilde{\tilde\tau}_{2n}\tilde{\tilde\tau}_{2n+2}}}
\frac{c_{2n}z^{2n+1}-c_{2n-1}z^{2n-1}}{c_0c_2\cdots c_{2n}}=
\frac{1}{\sqrt{c_{2n}}}(c_{2n}z^{2n+1}-c_{2n-1}z^{2n-1}).\nonumber
\eea
Then, also from appendix 1, in order to get
 $O \rightarrow \hat O$ in the correct form,
we compute the skew-orthonormal polynomials $\hat r_k$,
 with $\hat O \chi(z)=\hat r(z)$:
\bea
\hat r_{2n}(z)&=&\frac{1}{\sqrt{a_{2n}}}~r_{2n}(z)=\sqrt{\frac{c_{2n}}{a_{2n}}}
z^{2n}\nonumber\\
\hat r_{2n+1}(z)&=&\frac{\sum_0^{2n}b_i}{\sqrt{a_{2n}}}~r_{2n}(z)
+\sqrt{a_{2n}} r_{2n+1}(z)\nonumber\\
 &=&\sqrt{\frac{a_{2n}}{c_{2n}}}
\left(-c_{2n-1}z^{2n-1}+\frac{c_{2n}}{a_{2n}}(\sum_0^{2n}b_i)
 z^{2n}+c_{2n}z^{2n+1}\right).
\nonumber\\
\eea
From the coefficients of the polynomial $\hat r_k$, one reads off
the transformation matrix from orthonormal to skew-orthonormal
polynomials; it is given by the matrix $\hat O $, such that
 $\hat O\chi(z)=\hat r(z)$. Therefore
 $q(t,z)=\hat O(2t)p(2t,z)$ yields, after setting $t=0$,
\bea
 q_{2n}(0,z)&=&\sqrt{\frac{c_{2n}}{a_{2n}}}p_{2n}(0,z)\nonumber\\
 q_{2n+1}(0,z)&=&\sqrt{\frac{a_{2n}}{c_{2n}}}\nonumber\\
&&
\left(-c_{2n-1}p_{2n-1}(0,z)+\frac{c_{2n}}{a_{2n}}(\sum_0^{2n}b_i)
 p_{2n}(0,z)+c_{2n}p_{2n+1}(0,z)\right),\nonumber\\
\eea
confirming (6.2).

\medbreak

\section{Example 2: From Hermitean to symplectic matrix integrals}

\begin{proposition}
The matrix transformation
$$
\NR=f(L)M-\frac{f'+g}{2}(L),
$$
 maps the Toda lattice $\tau$-functions with $t$-dependent
  weight
  $$\rho_t(z)=e^{-V(z)+\sum_1^{\iy}t_iz^i},~V'=g/f $$
   (Hermitean matrix
integral) to the Pfaff lattice $\tau$-functions (Symplectic matrix
integral), with $t$-dependent weight
 $$\tilde \rho_t(z):=(\rho_{2t}(z)f(z))^{\frac{1}{2}}
 =e^{-\frac{1}{2}(V(z)-\log
f(z)-2\sum_1^{\iy}t_i z^i)}=:e^{-\tilde V(z)+\sum t_i z^i}.
 $$
 To be precise:


$$
\mbox{Toda lattice}\left\{\begin{tabular}{l}
$p_n(t,z) ~~\mbox{orthonormal polynomials in
 $z$ for the inner-product }$\\

\hspace{3.0cm}$\la\varphi,\psi\ra^{sy}_t=\int
 \varphi (z)\psi (z) \rho_t(z)dz $ \\  \\

$\mu_{ij}(t)=\la z^i,z^j\ra^{sy}_t,$ and $ m_n=
 (\mu_{ij})_{0\leq i,j\leq n-1},$\\
$\tau_n(t)=\det
m_n(t)=\displaystyle{\frac{1}{n!}\int_{\HR_n}e^{Tr(-V(X)+\sum
t_iX^i)}dX}$
\\
\end{tabular}
\right.
$$

$$\quad\left\downarrow\vbox to1.1in{\vss}\right.
\mbox{map $O(2t)$ such that}\
\left\{
\begin{array}{l}
  -\NR(2t)= O^{-1}(2t)JO^{\top -1}(2t)\\[6pt]
  O(2t) \mbox{ is lower-triangular } \\[6pt]
   O(2t)S(2t) \in \GR_{\Bk}
\end{array}
\right.
$$

$$
\mbox{Pfaff lattice}\left\{\begin{tabular}{l}
$q_n(t,z)=O (2t) p_n(2t,z) ~~\mbox{skew-orthonormal
  polynomials}$\\
\hspace{3.0cm}$\mbox{ in
$z$ for the skew-inner-product
 (weight $\tilde \rho_t$}$),
 \\

\\
 \hspace{1cm}$
 \la \varphi,\psi \ra_t^{sk}
 :=\la \varphi,\Bn_{2t}\psi \ra_{2t}^{sy}$\\  \\
$  \hspace{2.6cm}=\displaystyle{-\frac{1}{2}
 \int\!\int_{\BR^2}\{\varphi(z), \psi(z)\}
  \tilde\rho^2_t(z) dz}
 $\\ \\

$\tilde
\mu_{ij}(t)=\la z^i,z^j\ra^{sk}_t$, and
$\tilde m_n=\det(\tilde\mu_{ij})_{0\leq i,j\leq n-1}$\\
$\tilde\tau_{2n}(t)=pf(\tilde
m_{2n}(t))=\displaystyle{\frac{1}{(-2)^n n!}\int_{{\cal
T}_{2n}}e^{2Tr(-\tilde V(X)+\sum t_iX^i)}dX}.$
\\
\end{tabular}
\right.
$$
\end{proposition}

\bigbreak

\proof
Representing $d/dx$ as an integral operator
$$
\frac{d}{dx}\vp(x)=\int_{\BR}\delta(x-y)\vp'(y)dy=-\int_{\BR}\frac{\pl}{\pl
y}\delta(x-y)\vp(y)dy=\int_{\BR}\delta'(x-y)\vp(y)dy,
$$
compute
$$
\Bu_t=\Bn_t= \sqrt{\frac{f}{\rho_t}}\frac{d}{dz}
 \sqrt{ f \rho_t} ,
~~\mbox{so that ~$\Bn_tp(t,z)=\NR p(t,z) $;}
$$ remember $\NR$ from (5.2). Let it act on a function $\vp(x)$:
\begin{eqnarray*}
\Bu_t\vp(x)
&=&\left(\sqrt{\frac{f}{\rho_t}}\frac{d}{dx}\sqrt{f\rho_t}\right)\vp(x)\\
&=&\int_{\BR}\sqrt{\frac{f(x)}{\rho_t(x)}}\delta'(x-y)
\sqrt{f(y)\rho_t(y)}\vp(y)dy.
\end{eqnarray*}
Then
\begin{eqnarray*}
\la\vp,\psi\ra_t^{sk}&=&\la\vp,\Bu_{2t}\psi\ra_{2t}^{sy}
 =\la\vp,\Bn_{2t}\psi\ra_{2t}^{sy}\\
&=&\int\!\int_{\BR^2}\rho_{2t}(x)\vp(x)\sqrt{\frac{f(x)}
{\rho_{2t}(x)}}\delta'(x-y)\sqrt{f(y)\rho_{2t}(y)}
\psi(y)dx\,dy\\
&=&\int\!\int_{\BR^2}\sqrt{f(x)\rho_{2t}(x)}\vp(x)
\delta'(x-y)\sqrt{f(y)\rho_{2t}(y)}\psi(y)dx\,dy\\
&=&-\int\!\int_{\BR^2}\left(\frac{\pl}{\pl
x}\sqrt{f(x)\rho_{2t}(x)}\vp(x)\right)\delta(x-y)
\sqrt{f(y)\rho_{2t}(y)}\psi(y)dx\,dy\\
&=&-\int_{\BR}\left(\frac{\pl}{\pl
x}\sqrt{f(x)\rho_{2t}(x)}\vp(x)\right)\sqrt{f(x)\rho_{2t}(x)}\psi(x)dx\\
&=&-\frac{1}{2}\int_{\BR}\left(\frac{\pl}{\pl
x}\sqrt{f(x)\rho_{2t}(x)}\vp(x)\right)\sqrt{f(x)\rho_{2t}(x)}\psi(x)dx\\
& &\quad
+\frac{1}{2}\int_{\BR}\sqrt{f(x)\rho_{2t}(x)}\vp(x)
 \left(\frac{\pl}{\pl x}\sqrt{f(x)\rho_{2t}(x)}\psi(x)\right)dx\\
&=&-\frac{1}{2}\int_{\BR}\{\sqrt{f(x)\rho_{2t}(x)}\vp(x),
 \sqrt{f(x)\rho_{2t}(x)}\psi(x)\}dx\\
&=&-\frac{1}{2}\int_{\BR}\{\vp(x),\psi(x)\}\tilde\rho_0^2(x)
 e^{2\sum_1^{\iy}t_i x^i}dx,
\end{eqnarray*}
using the notation in the statement of this proposition. Setting
$\tilde \rho(x)=e^{-\tilde V(x)}$, with $\tilde
V(x)=\frac{1}{2}(V(x)-\log f)$
\begin{eqnarray*}
\la x^i,x^j\ra^{sk}&=&-\frac{1}{2}\int_{\BR}\{x^i,x^j\}
\tilde\rho^2(x)
 e^{2\sum_1^{\iy}t_i x^i}dx\\
&=&-\frac{1}{2}\int_{\BR}\{x^i,x^j\} e^{-2(\tilde V(x)-\sum
t_ix^i)}dx,
\end{eqnarray*}
and so
$$
\tau_{2n}(t)=pf (\tilde m_{2n}(t))
=\frac{1}{(-2)^n n!}\int_{\TR_{2n}}
e^{2Tr(-\tilde V(x)+\sum t_ix^i)}dx.
$$

\bigbreak

\noindent{\bf The map $O^{-1}$ for the
classical orthogonal polynomials at $t=0$}: {\em Then, the matrix
$O$, mapping orthonormal $p_k$ into skew-orthonormal polynomials
$q_k$, is given by a lower-triangular three-step relation :
\bea
p_{2n}(0,z)&=&
-c_{2n-1}\sqrt{\frac{a_{2n-2}}{c_{2n-2}}}q_{2n-2}(0,z)
+\sqrt{a_{2n}c_{2n}}~q_{2n}(0,z)\nonumber\\
 p_{2n+1}(0,z)&=& -c_{2n}\sqrt{\frac{a_{2n-2}}{c_{2n-2}}}
 q_{2n-2}(0,z)-(\sum_0^{2n} b_i)\sqrt{\frac{c_{2n}}{a_{2n}}}
 q_{2n}(0,z)+\sqrt{\frac{c_{2n}}{a_{2n}}}q_{2n+1}(0,z),\nonumber\\
 \eea
where the $a_i$ and $b_i$ are the entries in the tridiagonal matrix
defining the orthonormal polynomials, and the $c_i$ the entries in
the skew-symmetric matrix.}

In this case, we need to perform the following skew-Borel
decomposition at $t=0$,
$$-\UR=-\NR=O^{-1}JO^{\top -1},
$$
where $\NR$ is the matrix (6.3). Here again, in order to find $O$,
we use the recipe given in theorem 4.1, namely writing down the
corresponding skew-orthogonal polynomials (0.17), but where the
 $\mu_{ij}$ are the entries of $-\UR=-\NR$:
consider the pfaffians of the bordered matrices (0.17); they have
leading term
$$\tilde{\tilde\tau}_{2n}= \prod_0^{n-1} c_{2j}.$$
 Then one computes
\bea
r_{2n}&=&\frac{1}{\sqrt{\tilde{\tilde\tau}_{2n}\tilde{\tilde\tau}_{2n+2}}}\sum_{i=0}^n
 z^{2n-2i}\left( \prod_0^{n-i-1} c_{2j} \right)
 \left( \prod_0^{i-1} c_{2n-2j-1} \right)\nonumber\\
r_{2n+1}&=&\frac{1}{\sqrt{\tilde{\tilde\tau}_{2n}\tilde{\tilde\tau}_{2n+2}}}\left(z^{2n+1}
\prod_0^{n-1} c_{2j}+\sum_{i=1}^n
 z^{2n-2i}\left( \prod_0^{n-i-1} c_{2j} \right)
 \left( \prod_0^{i-1} c_{2n-2j-1} \right)\right)\nonumber\\
 \eea
 with
 $$\sqrt{\tilde{\tilde\tau}_{2n}\tilde{\tilde\tau}_{2n+2}}=
c_0 c_2...c_{2n-2}\sqrt{c_{2n}}
~~~
 \sqrt{\tilde{\tilde\tau}_{0}\tilde{\tilde\tau}_{2}}=\sqrt{c_{0}}
.$$
Setting
$$
D:=\diag
(\sqrt{\tilde{\tilde\tau}_0\tilde{\tilde\tau}_2},\sqrt{\tilde{\tilde
\tau}_0\tilde{\tilde\tau}_2},\sqrt{\tilde{\tilde\tau}_2\tilde{\tilde
\tau}_4},
\sqrt{\tilde{\tilde\tau}_2\tilde{\tilde\tau}_4},...)
,$$ the matrix $O$ is the set of coefficients of the polynomials
 above, i.e.,

\bea O&=&D^{-1}\pmatrix{1&0&0&0&0&0&0&0\cr 0&1&0&0&0&0&0&0\cr  c_1&0&%
 {\rm c_0}&0&0&0&0&0\cr {\rm c_2}&0&0&{\rm c_0}&0&0&0&0\cr {\rm %
 c_1}\,{\rm c_3}&0&{\rm c_0}\,{\rm c_3}&0&c_0 c_2&0&0&0%
 \cr {\rm c_1}\,{\rm c_4}&0&{\rm c_0}\,{\rm c_4}&0&0&{\rm c_0}
 \,{\rm c_2}&0%
 &0\cr {\rm c_1}\,{\rm c_3}\,{\rm c_5}&0&{\rm c_0}\,{\rm c_3}
 \,{\rm c_5%
 }&0&{\rm c_0}\,{\rm c_2}\,{\rm c_5}&0&{\rm c_0}\,{\rm c_2}\,{\rm c_4}
 &0\cr {\rm c_1}\,{\rm c_3}\,{\rm c_6}&0&{\rm c_0}\,{\rm c_3}\,{\rm c_6}&0&%
 {\rm c_0}\,{\rm c_2}\,{\rm c_6}&0&0&{\rm c_0}\,{\rm c_2}\,{\rm c_4}%
 \cr &&&&&&&&\ddots
 }\nonumber\\
 &=:&D^{-1}R
 \eea
As before, in order to get the skew-symmetric polynomials in the
right form, from the orthogonal ones, one needs to multiply to the
left with the matrix $E$, defined in (8.2) in the appendix:
 \be
\hat O=EO=ED^{-1}R,
\ee
and so,
 \be
\hat O^{-1}=R^{-1}DE^{-1};
\ee
it turns out the matrix $\hat O$ is complicated, but its inverse is
simple. Namely, compute
$$R^{-1}= \pmatrix{1&0&0&0&0&0&0&0\cr 0&1&0&0&0&0&0&0\cr
 -{{{ c_1}%
 }\over{{ c_0}}}&0&{{1}\over{{\rm c_0}}}&0&0&0&0&0\cr
  -{{{\rm c_2%
 }}\over{{\rm c_0}}}&0&0&{{1}\over{{\rm c_0}}}&0&0&0&0\cr
  0&0&-{{%
 {\rm c_3}}\over{{\rm c_0}\,{\rm c_2}}}&0&{{1}\over{{\rm c_0}\,
 {\rm c_2}}}%
 &0&0&0\cr 0&0&-{{{\rm c_4}}\over{{\rm c_0}\,{\rm c_2}}}&0&0&{{1%
 }\over{{\rm c_0}\,{\rm c_2}}}&0&0\cr 0&0&0&0&-{{{\rm c_5}}\over{%
 {\rm c_0}\,{\rm c_2}\,{\rm c_4}}}&0&{{1}\over{{\rm c_0}\,{\rm c_2}\,{\rm %
 c_4}}}&0\cr 0&0&0&0&-{{{\rm c_6}}\over{{\rm c_0}\,{\rm c_2}\,{\rm c_4%
 }}}&0&0&{{1}\over{{\rm c_0}\,{\rm c_2}\,{\rm c_4}}}\cr
 \cr&&&&&&&&\ddots }  ,$$
 and
\be
E^{-1}
=\left(
\begin{array}{c@{}c@{}cc}
 \boxed{\begin{array}{cc} \al_0 & 0 \\ -\beta_0 & \frac{1}{\al_0}
 \end{array}} &&&0 \\
 & \boxed{\begin{array}{cc}  \al_2 & 0 \\ -\beta_2&\frac{1}{\al_2} \end{array}} &&\\
 && \boxed{\begin{array}{cc} \al_4&0 \\ -\beta_4&\frac{1}{\al_4} \end{array}} & \\
 0& && \ddots
 \end{array}
 \right),
\ee
with $\al_{2n}$ and $\beta_{2n}$ as in (8.5). Carrying out the
multiplication (7.5) leads to the matrix $\hat O^{-1}$, with a few
non-zero bands, yielding the map (7.1).

\section{Appendix 1: Free parameter in the skew-Borel decomposition}

If the Borel decomposition of $-H=O^{-1}JO^{\top -1}$ is given by a
matrix $O \in \GR_{\Bk}$, with the diagonal part of $O$ being
\be
(O)_0=\left(
\begin{array}{c@{}c@{}cc}
 \boxed{\begin{array}{cc} \sg_0&0 \\ 0&\sg_0 \end{array}} && &0 \\
 & \boxed{\begin{array}{cc} \sg_2&0 \\ 0&\sg_2 \end{array}} &&\\
 && \boxed{\begin{array}{cc} \sg_4&0 \\ 0&\sg_4 \end{array}} & \\
 0&&& \ddots
 \end{array}
 \right),
\ee
then the new matrix
\be
\hat O:
=\left(
\begin{array}{c@{}c@{}cc}
 \boxed{\begin{array}{cc} 1/\al_0 & 0 \\ \beta_0 & \al_0
 \end{array}} &&&0 \\
 & \boxed{\begin{array}{cc}  1/\al_2 & 0 \\ \beta_2&\al_2 \end{array}} &&\\
 && \boxed{\begin{array}{cc} 1/\al_4&0 \\ \beta_4&\al_4 \end{array}} & \\
 0& && \ddots
 \end{array}
 \right)O=:EO,
\ee
with free parameters $\al_{2n},~\beta_{2n}$, is a solution of the
Borel decomposition $-H=\hat O^{-1}J\hat O^{\top
-1}$, as well. The diagonal part of $\hat O$ consists of $2\times 2$ blocks
$$
\pmatrix{1/\al_{2n}&0\cr
 \beta_{2n} & \al_{2n}}
\pmatrix{\sg_{2n}&0\cr
 0 & \sg_{2n}}=
\pmatrix{\sg_{2n}/\al_{2n}&0\cr
 \beta_{2n}\sg_{2n} & \al_{2n}\sg_{2n}}.
$$
Imposing the condition that
$$
q(z)=\hat O p(z), ~\mbox{with}~ p_k(z)=\sum^k_{i=0}p_{ki}z^i
$$
has the required form, i.e., the same leading term for $q_{2n}$ and
$q_{2n+1}$ and no $z^{2n}$-term in $q_{2n+1}$,
\bea
q_{2n}(z)&=&q_{2n,2n}z^{2n}+\cdots\nonumber\\
 q_{2n+1}(z)&=&q_{2n,2n}z^{2n+1}+q_{2n,2n-1}z^{2n-1}+\cdots
\eea
implies
$$
\frac{\sg_{2n}}{\al_{2n}}p_{2n,2n}=\sg_{2n} \al_{2n}p_{2n+1,2n+1}
$$
$$
 \sg_{2n}\beta_{2n}p_{2n,2n}+\sg_{2n} \al_{2n}p_{2n+1,2n}=0
$$
yielding, upon using the explicit form of the coefficients
$p_{k\ell}$ of the polynomials $p_k$, associated with three step
relations (see next lemma),
\bea
\al_{2n}^2&=&\frac{p_{2n,2n}}{p_{2n+1,2n+1}}=a_{2n}\nonumber\\
 \frac{\beta_{2n}}{\al_{2n}}&=&
    -\frac{p_{2n+1,2n}}{p_{2n,2n}}=\frac{\sum_0^{2n} b_i}{a_{2n}}.
\eea
Hence
\be
 \al_{2n}=\sqrt{a_{2n}}~~~\mbox{and}~~~\beta_{2n}=
 \frac{1}{\sqrt{a_{2n}}}
 \sum_0^{2n} b_i.
 \ee

So, if
$$
r(z)=O\chi(z),
$$
then
$$
\hat r (z):=\hat O\chi(z)=\left(
\begin{array}{c@{}c@{}cc}
 \boxed{\begin{array}{cc} 1/\al_0 & 0 \\ \beta_0 & \al_0
 \end{array}} &&&0 \\
 & \boxed{\begin{array}{cc}  1/\al_2 & 0 \\ \beta_2&\al_2 \end{array}} &&\\
 && \boxed{\begin{array}{cc} 1/\al_4&0 \\ \beta_4&\al_4 \end{array}} & \\
 0& && \ddots
 \end{array}
 \right)  r(z)=E r(z),
 $$
 and thus
 \bea
 \hat r_{2n}(z)&=&\frac{1}{\sqrt{a_{2n}}}r_{2n}(z)\\
 \hat r_{2n+1}(z)&=&\frac{\sum_0^{2n} b_i}{\sqrt{a_{2n}}}
r_{2n}+\sqrt{a_{2n}}r_{2n+1}(z)
\eea

\begin{lemma} A sequence of polynomials $p_n(z)=\sum^n_{i=0}p_{ni}z^i$ of
degree $n$ satisfying three-step recursion relation\footnote{with
$a_{-1}=0$.}
\be
zp_n=a_{n-1}p_{n-1}+b_np_n+a_np_{n+1},\quad n=0,1,\dots,
\ee
has the form
$$
p_{n+1}(z)=\frac{p_{n,n}}{a_n}\left(z^{n+1}-
(\sum_0^{n}b_i)z^n+\cdots\right).
$$
\end{lemma}

\proof Equating the $z^{n+1}$ and $z^n$ coefficients of (8.8) divided by
$p_{n,n}$ yields
$$
\frac{p_{n+1,n+1}}{p_{n,n}}=\frac{1}{a_n}
$$
and
$$
\frac{p_{n,n-1}}{p_{n,n}}=a_n\frac{p_{n+1,n}}{p_{n,n}}+b_n.
$$
Combining both equations leads to
$$
a_n\frac{p_{n+1,n}}{p_{n,n}}-a_{n-1}\frac{p_{n,n-1}}{p_{n-1,n-1}}=b_n,
$$
yielding
$$
a_n\frac{p_{n+1,n}}{p_{n,n}}=-\sum_0^nb_i,\quad \mbox{using $a_{-1}=0$}.
$$
\qed

\section{Appendix 2: Simultaneous (skew) - symmetrization of $L$ and $\NR$.}

{\em For the classical polynomials, the matrices $L$ and $\NR$ can
be simultaneously symmetrized and skew-symmetrized}.

We sketch the proof of this statement, which has been established
by us in \cite{AvM2}. Given the monic orthogonal polynomials
$\tilde p_n$ with respect to the weight $\rho$, with
$\rho'/\rho=-g/f$, we have that the operators $z$ and
 $$
 \Bn=\sqrt{\frac{f}{\rho}} \frac{d}{dz} \sqrt{f\rho}
 =f \frac{d}{dz} +\frac{f'-g}{2}.
 $$
acting on the polynomials $\tilde p_n$'s have the following form:
\bea
z\tilde p_n&=&a_{n-1}^2\tilde p_{n-1}+b_n\tilde p_n+\tilde p_{n+1}
 \nonumber\\
 \Bn\tilde p_n&=&
 ...-\gamma_n \tilde p_{n+1},
 \eea
 in view of the fact that for the classical orthogonal
 polynomials\footnote{with respective weights $\rho=e^{-z^2},~
 \rho=e^{-z}z^{\al},~
\rho=(1-z)^{\al}(1+z)^{\beta}$.},

$\left\{
\noindent\begin{tabular}{l l }
Hermite: & $\Bn=\frac{d}{dz}-z $\\  Laguerre: &
$
\Bn=z\frac{d}{dz}-
\frac{1}{2}(z-\alpha -1)$ \\  Jacobi: & $\Bn=(1-z^2)\frac{d}{dz}
-\frac{1}{2}((\alpha +\beta
+2)z+(\alpha -\beta)).$\\
\end{tabular}\right.
$

For the orthonormal polynomials, the matrices $L$ and $-\NR$ are
symmetric and skew-symmetric respectively. Therefore the right hand
side of these expressions must have the form:
\bean
z\tilde p_n&=&a_{n-1}^2\tilde p_{n-1}+b_n\tilde p_n+\tilde p_{n+1}
 \\
 \Bn\tilde p_n&=&
 a^2_{n-1}\gamma_{n-1}\tilde p_{n-1}-\gamma_n \tilde p_{n+1}.
 \eean
Therefore, upon rescaling the $\tilde p_n$'s, to make them
orthonormal, we have
\bean
z p_n&=&(Lp)_n =a_{n-1} p_{n-1}+b_n p_n+ a_n p_{n+1}
 \\
 \Bn  p_n&=& (\NR p)_n=
 a_{n-1}\gamma_{n-1}\tilde p_{n-1}-a_n\gamma_n \tilde p_{n+1},
 \eean
from which it follows that $$-\NR=\left[\begin{array}{ccccc} 0&c_0&
& & \\
-c_0&0&c_1& & \\
 &-c_1&0&\ddots&\\
 & &\ddots& &
\end{array}\right],~~
\mbox{with}~~ c_n=-a_n \gamma_n~,
$$
where $-\gamma_n$ is the leading term in the expression (9.1).

\section{Appendix 3: Proof of Lemma 3.4}

For future use, consider the first order differential operators
 \be
 \eta(t,z)=\sum^{\iy}_{j=1}\frac{z^{-j}}{j}\frac{\pl}{\pl
 t_j}\quad\mbox{and}\quad
 B(z)=-\frac{\pl}{\pl z}+\sum^{\iy}_{j=1}z^{-j-1}\frac{\pl}{\pl t_j}
 \ee
 having the property
 \be
 B(z)e^{-\eta(z)}f(t)=B(z)f(t-[z^{-1}])=0.
 \ee

\begin{lemma} Consider an arbitrary function $\vp(t,z)$ depending on
$t\in\BC^{\iy}$, $z\in\BC$, having the asymptotics
$\vp(t,z)=1+O(\frac{1}{z})$ for $z\nearrow\iy$ and
 satisfying the functional
relation
\be
\frac{\vp(t-[z^{-1}_2],z_1)}{\vp(t,z_1)}=\frac{\vp(t-[z^{-1}_1],z_2)}
{\vp(t,z_2)}, \quad t\in\BC^{\iy},z\in\BC.
\ee
Then there exists a function $\tau(t)$ such that
\be
 \vp(t,z)=\frac{\tau(t-[z^{-1}])}{\tau(t)}.
\ee
\end{lemma}

\proof
Applying $B_1:=B(z_1)$ to the logarithm of (10.3) and using
 (10.1) and (10.3) yields
 \begin{eqnarray*}
 (e^{-\eta(z_2)}-1)B_1\log\vp(t;z_1)&=&-B_1\log\vp(t,z_2)\\
 &=&\sum^{\iy}_{j=1}z_1^{-j-1}\frac{\pl}{\pl t_j}\log\vp(t,z_2),
 \end{eqnarray*}
 which, upon setting
 $$
 f_j(t)=\mbox{\,Res}_{z_1=\iy}z_1^jB_1\log\vp(t,z_2)
, $$
 yields termwise in $z_1$,
 \be
 (e^{-\eta(z_2)}-1)f_j(t)=-\frac{\pl}{\pl t_j}\log\vp(t,z_2).
 \ee
 Acting with $\frac{\pl }{\pl t_i}$ on the latter expression
 and with $ \frac{\pl }{\pl t_j}$
 on the same expression with $j$ replaced by $i$, and subtracting\footnote{It
 is obvious that
 $\left[\frac{\pl}{\pl t_i},e^{-\eta(z)}\right]=0$.}, one finds
 $$
 (e^{-\eta(z_2)}-1)\left(\frac{\pl f_i}{\pl t_j}-
 \frac{\pl f_j}{\pl t_i}\right)=0,
 $$
 yielding
 $$
 \frac{\pl f_i}{\pl t_j}-\frac{\pl f_j}{\pl t_i}=0;
 $$
 the constant vanishes, because $\frac{\pl f_i}{\pl t_j}$ never contains
 constant terms.

 Therefore there exists a function
 $\log\tau(t_1,t_2,...)$ such that
 $$
 -\frac{\pl }{\pl t_j}\log\tau=f_j(t)=\mbox{\,Res}_{z=\iy}z^jB\log\vp
 $$
 and hence, using (10.5)
 $$
 \frac{\pl }{\pl t_j}\log\vp(t,z)=(e^{-\eta(z)}-1)
 \frac{\pl }{\pl
 t_j}\log\tau
 $$
 or, what is the same,
 $$
 \frac{\pl }{\pl t_j}(\log\vp-(e^{-\eta}-1)\log\tau)=0,
 $$
 from which it follows that
 $$
 \log\vp-(e^{-\eta}-1)\log\tau=-\sum_1^{\iy}\frac{b_i}{i}z^{-i}
 $$
 is, at worst, a holomorphic series in $z^{-1}$ with constant coefficients,
 which we call $-b_i/i$. Hence
 \begin{eqnarray*}
 \vp(t,z)&=&\frac{\tau(t-[z^{-1}]e^{-\sum_1^{\iy}\frac{b_i}{i}
 z^{-i}}}{\tau(t)}\\
 &=&\frac{\tau(t-[z^{-1}])e^{\sum_1^{\iy}b_i(t_i-\frac{z^{-i}}{i})}}
 {\tau(t)e^{\sum_1^{\iy}b_it_i}},
 \end{eqnarray*}
 i.e.
 $$
 \vp(t,z)=\frac{\tilde\tau(t-[z^{-1}])}{\tilde\tau(t)},
 $$
 where
 $$
 \tilde\tau=\tau(t)e^{\sum_1^{\iy}b_it_i}.
 $$
 \qed


\begin{thebibliography}{99}

\bibitem{AvM1} M. Adler and P.~ van Moerbeke: {\em Completely
integrable system, Euclidean Lie algebras and curves}, Advances in
Math. 38, 267-317 (1980) .


\bibitem{AvM2} M.~ Adler and P.~ van Moerbeke: {\em Matrix
integrals, Toda symmetries, Virasoro constraints and orthogonal
polynomials} Duke Math.J., {\bf 80} (3), 863--911 (1995)


\bibitem{AvM3} M. Adler and P. van Moerbeke: {\em Group factorization,
moment matrices and 2-Toda lattices},   Intern. Math. Research Notices,
{\bf 12}, 555-572 (1997)

\bibitem{AvM4} M. Adler and P. van Moerbeke: {\em The spectrum of symmetric random
matrices and the Pfaff Lattice}, Annals of Math.,
(2000) (to appear).








 \bibitem{AHV} M.~Adler, ~E.~Horozov and P.~ van Moerbeke :
{\em The Pfaff lattice and skew-orthogonal polynomials},
 Int. Math. Res. Notices, {\bf 11} 569-588 (1999).


 \bibitem{ASV} M.~Adler, T.~Shiota and P.~ van Moerbeke :
{\em The Pfaffian $\tau$-functions } (to appear).


\bibitem{BN} E.~Br\'ezin, H.~Neuberger : {\em
Multicritical points of unoriented random surfaces}, Nuclear
Physics {\bf B 350}, 513-553 (1991).

\bibitem{Dickey} L. Dickey: Soliton equations and integrable
systems, World Scientific (1991).





\bibitem{Date} E. Date, M. Jimbo, M. Kashiwara, T. Miwa: {\em
Transformation groups for soliton equations}, In: Proc.\ RIMS
Symp.\ Nonlinear integrable systems --- Classical and quantum
theory (Kyoto 1981), pp.\ 39--119.\ Singapore : World Scientific
1983.











\bibitem{For} P. Forrester: Random matrices, Cambridge University press.

\bibitem{Kac} V.G. Kac: Infinite dimensional Lie algebras, 3rd
edition, Cambridge University press.


\bibitem{KvdL} V.G. Kac and J. van de Leur: {\em
 The geometry of spinors and the multicomponent
 BKP and DKP hierarchies} in "The bispectral problem
 (Montreal PQ, 1997)", CRM Proc. Lecture notes {\bf 14},
 AMS, Providence, 159-202 (1998).



\bibitem{M} M.L. Mehta: Random matrices, 2nd ed.\
Boston: Acad.\ Press, 1991

\bibitem{AFNV} M.~Adler, P.J.~Forrester,
T.~Nagao and P.~van Moerbeke:
{\em
Classical skew orthogonal polynomials and random
matrices},
Journal of Statistical Physics, 2000




\bibitem{RS} A.G.~Reyman and M.A. Semenov-Tian-Shansky : {\em
 Reduction of Hamiltonian systems, affine Lie algebras and
 Lax equations }, Inv. Math., {\bf 54}, 81-100 (1979).

\bibitem{vdL} J. van de Leur: {\em
Matrix Integrals and Geometry of Spinors }
 (solv-int/9909028)






\end{thebibliography}
\end{document}